\documentclass[prb,amsmath,amsfont,amssymb,twocolumn,showpacs]{revtex4-1}
\usepackage{graphicx}
\usepackage{color}
\usepackage[english]{babel}
\usepackage{amsmath}

\renewcommand{\phi}{\varphi}
\renewcommand{\epsilon}{\varepsilon}

\newcommand\op[1]{\hat{#1}} 
\newcommand\vekk[1]{\mathbf{#1}} 
\newcommand\matr[1]{\mathbf{#1}} 
\newcommand\norm[1]{\lvert #1 \rvert}
\newcommand\mean[1]{\langle #1 \rangle}

\newcommand\kv{\vekk{k}}
\newcommand\kpv{\vekk{k}'}

\newcommand\rv{\vekk{r}}

\newcommand\qv{\vekk{q}}
\newcommand\qcv{\vekk{q}_c}

\newcommand\epsHF{\epsilon}
\newcommand\kfo{k_F^{0}}
\newcommand\Efo{E_F^{0}}

\DeclareMathAlphabet{\mathcalligra}{T1}{calligra}{m}{n}
\newcommand\ordpar{\mathcalligra{h}}
\newcommand\erfc{{\rm erfc}}

\newcommand\HMF{\op{H}_{{\rm MF}}}
\newcommand\Hop{\op{H}}

\newcommand\co[1]{\op{c}_{#1}}
\newcommand\cd[1]{{\op{c}_{#1}}^{\dagger}}
\newcommand\go[1]{\op{\gamma}_{#1}}
\newcommand\gd[1]{{\op{\gamma}_{#1}}^{\dagger}}

\begin{document}

\title{Properties of the density-wave phase of a two-dimensional dipolar fermi gas}

\author{J. K. Block}
\email{jkblock@phys.au.dk}
\affiliation{Department of Physics and Astronomy, Aarhus University, DK-8000 Aarhus C, Denmark}

\author{G.\ M.\ Bruun}
\affiliation{Department of Physics and Astronomy, Aarhus University, DK-8000 Aarhus C, Denmark}

\date{\today}

\begin{abstract}
The rapid progress in the production and cooling of molecular gases indicates that experimental studies of quantum gases with a strong dipolar interaction is 
soon within reach. Dipolar gases are predicted to exhibit very rich physics including quantum liquid crystal phases such as density-waves as 
well as superfluid phases, both of which play an important role for our understanding of strongly correlated systems.
Here, we investigate the zero temperature properties of the density-wave phase of a two-dimensional (2D) system of fermonic dipoles using a conserving Hartree-Fock theory.
We calculate the amplitude of the density waves as a function of the dipole moment and orientation with respect to the 2D plane.
The stripes give rise to a 1D Brillouin zone structure, and the corresponding quasiparticle spectrum is shown to have gapped as well as gapless regions around
the Fermi surface. As a result, the system remains compressible in the density-wave phase, and it collapses for strong attraction. We show that
the  density-waves has clear signatures in the momentum distribution and in the momentum correlations. Both can be measured
in time-of-flight experiments. Finally, we discuss how the striped phase can be realised with experimentally available systems.
\end{abstract}


\maketitle

\section{Introduction}
The investigation of ultracold atomic gases has produced several breakthrough results in
the last two decades~\cite{Bloch2008,Giorgini2008a}. Atomic gases are used to simulate many-body systems without the
presence of disorder, intricate band structures etc., which significantly complicates the
understanding of conventional condensed matter systems. One limitation is that
the atom-atom interaction is typically short range and isotropic ($s$-wave), whereas the order parameters
in nature often exhibit richer $p$- and $d$-wave symmetries. The impressive progress in the
production of cold gases consisting of fermionic hetero-nuclear molecules with an electric dipole moment~\cite{Ni2010,Deh2010,Heo2012,Wu2012,Schulze2013,Tung2013,Repp2013}
promises to remove this limitation, since the dipole-dipole interaction is long-range and anisotropic
with both repulsive and attractive parts~\cite{Lahaye2009}.
The attractive head-to-tail part of the dipole-dipole interaction can lead to severe losses via chemical reactions, which however can be
 suppressed by orders of magnitude by confining the dipoles to low dimensional geometries~\cite{DeMiranda2011,Chotia2012},
 or by using molecules which are chemically stable such as ${}^{23}\textup{Na}^{40}\textup{K}$~\cite{Wu2012,Schulze2013} or ${}^{40}\textup{K}^{133}\textup{Cs}$.
 Dipolar gases are predicted to exhibit a wealth of new phases in 2D, including $p$-wave superfluids~\cite{Bruun2008,Cooper2009} as well as quantum
 liquid crystals such as nematic~\cite{Fregoso2009}, density-wave (smectic)~\cite{Sun2010,Yamaguchi2010,Zinner2011,Babadi2011,Sieberer2011,Parish2012,Block2012,Marchetti2013}
 and hexatic phases~\cite{Bruun2014,Lechner2014}. The presence of both superfluid and liquid crystal order occurs in several strongly correlated systems, 
 and it plays a central role in the physics of the cuprate and pnictide superconductors~\cite{Kivelson1998,Fradkin2010}.

We analyse in this paper the zero temperature ($T=0$) properties of the density-wave phase of a 2D gas of
fermionic dipoles aligned by an external field. In this phase, the dipoles form density waves (stripes) in order to minimise the repulsive side-by-side part of the interaction.
Several groups have predicted a 2D dipolar gas to form such a striped phase for large dipole
 moments~\cite{Sun2010,Yamaguchi2010,Zinner2011,Babadi2011,Sieberer2011,Parish2012,Block2012}.

Partitioning momentum space into one dimensional Brillouin zones,
we develop a conserving Hartree-Fock approximation (HFA)~\cite{Baym1961}, which is shown to recover previous results for the critical coupling strength
for the onset of stripe formation. The resulting equations are solved self-consistently, and we calculate the amplitude of the stripes as a function
of the dipole moment and its orientation with respect to the 2D plane.
We then calculate the quasi-particle spectrum and show that the 1D Brillouin zone structure gives rise to a Fermi surface with gapless as well as gapped regions.
As a result, the system remains compressible, and the formation of stripes does not stabilise the system against collapse for large dipole attraction.
The presence of stripes is demonstrated to have clear signatures in the momentum distribution
and to give rise to characteristic momentum correlations, both of which can be measured in time-of-flight experiments. Finally, we show
how the effects described in this paper can be observed with experimentally available dipolar gases.

\begin{figure}[htb]
 \centering
\def\svgwidth{\columnwidth}
\begingroup%
  \makeatletter%
  \providecommand\color[2][]{%
    \errmessage{(Inkscape) Color is used for the text in Inkscape, but the package 'color.sty' is not loaded}%
    \renewcommand\color[2][]{}%
  }%
  \providecommand\transparent[1]{%
    \errmessage{(Inkscape) Transparency is used (non-zero) for the text in Inkscape, but the package 'transparent.sty' is not loaded}%
    \renewcommand\transparent[1]{}%
  }%
  \providecommand\rotatebox[2]{#2}%
  \ifx\svgwidth\undefined%
    \setlength{\unitlength}{865.95bp}%
    \ifx\svgscale\undefined%
      \relax%
    \else%
      \setlength{\unitlength}{\unitlength * \real{\svgscale}}%
    \fi%
  \else%
    \setlength{\unitlength}{\svgwidth}%
  \fi%
  \global\let\svgwidth\undefined%
  \global\let\svgscale\undefined%
  \makeatother%
  \begin{picture}(1,0.894471)%
    \put(0,0){\includegraphics[width=\unitlength]{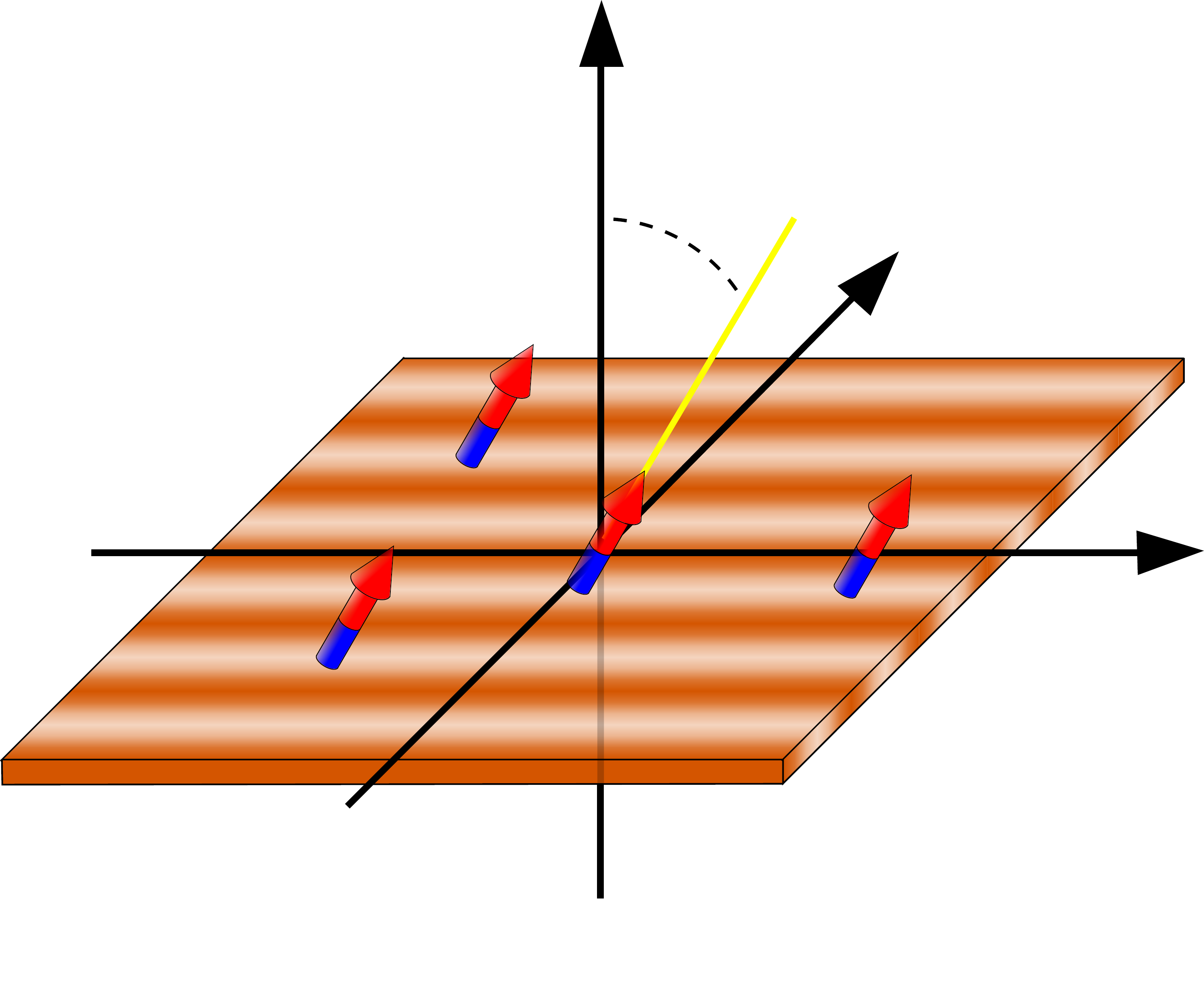}}%
    \put(0.6497318,0.67625213){\color[rgb]{0,0,0}\makebox(0,0)[lb]{\smash{$\mathbf{E}$}}}%
    \put(0.87857892,0.31793926){\color[rgb]{0,0,0}\makebox(0,0)[lb]{\smash{x}}}%
    \put(0.75073359,0.5701746){\color[rgb]{0,0,0}\makebox(0,0)[lb]{\smash{y}}}%
    \put(0.43284787,0.736028){\color[rgb]{0,0,0}\makebox(0,0)[lb]{\smash{z}}}%
    \put(0.55227251,0.65424897){\color[rgb]{0,0,0}\makebox(0,0)[lb]{\smash{$\theta$}}}%
  \end{picture}%
\endgroup%
 \caption{(colour online). The dipoles are confined in the $xy$-plane, and they are aligned by an external electrical field $\mathbf{E}$ forming the angle $\theta$ with respect to the $z$-axis. The density waves are along the $x$-axis which is defined by the projection of the field $\mathbf{E}$ onto the plane.}
 \label{fig:system}
\end{figure}
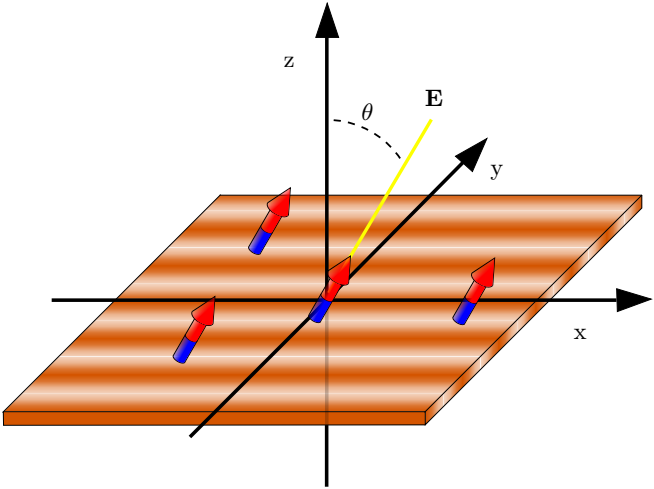

\section{System}
We consider identical fermionic dipoles of mass $m$ moving in a 2D layer defining the $xy$-plane at $T=0$.
The dipole moment $\mathbf{p}$ of the fermions is aligned
forming the angle $\theta$ with respect to the normal of the layer ($z$-axis) with its projection onto the planes defining the $x$-axis,
see figure~\ref{fig:system}.
We assume that the layer is formed by a deep 1D optical lattice so that the dipoles reside in the lowest harmonic oscillator level
$\phi(z)=\exp(-z^2/2w^2)\pi^{-1/4}w^{-1/2}$ in the z-direction with $w$ the width of the layer.
 We neglect any trapping potential in the $xy$-plane so that the transverse states are labelled by the momentum $\mathbf{k}=(k_x,k_y)$ (we
take $\hbar=k_B=1$).

The Hamiltonian of the system is
\begin{equation}
\Hop=\sum_{\kv}\frac{k^2}{2m}\cd{\kv}\co{\kv}+\frac{1}{2A}\sum_{\kv,\kpv,\qv}V(\qv)\cd{\kv+\qv}\cd{\kpv-\qv}\co{\kpv}\co{\kv},
 \label{eq:Hamiltonian}
\end{equation}
where $\hat c_{\kv}$ removes a dipole with momentum $\kv$. The interaction between two dipoles separated by $\mathbf{r}$ is
 $V_{\rm 3D}(\mathbf{r})=D^2[1-3\cos^2(\theta_r)]/r^3$
where $\theta_r$ is the angle between $\mathbf{r}$ and the dipole moment $\mathbf{p}$, and $D^2=p^2/4\pi\epsilon_0$
for electric dipoles. The effective interaction $V(\qv)$ in (\ref{eq:Hamiltonian}) is obtained by integrating the interaction $V_{\rm 3D}(\mathbf{r})$ over the
 Gaussians $\phi(z)^2$, which yields~\cite{Fischer2006}
\begin{equation}
V(\qv)=\pi D^2\bigg [\frac{8}{3w\sqrt{2\pi}} P_2(\cos\theta)-2\xi(\theta,\phi)F(q)\bigg ].
\label{eq:V2Dq}
\end{equation}
Here, $P_2(x)=(3x^2-1)/2$ is the second Legendre polynomial,
$F(q)= q\exp[(qw)^2/2]\erfc(qw/\sqrt{2})$ and $\xi(\theta,\phi)=\cos(\theta)^2-\sin(\theta)^2\cos(\phi)^2$.
The constant term in equation (\ref{eq:V2Dq}) corresponds to a contact interaction which plays no role here, since we
consider identical fermions. We characterise the strength of the interaction by the ratio of the typical interaction and kinetic energy,
\begin{equation}
 \label{eq:gdef}
 g=\frac{4mD^2\kfo}{3\pi\hbar^2},
\end{equation}
 where $\kfo=\sqrt{4\pi \rho_0}$ is defined from the areal density $\rho_0$. We likewise define the Fermi energy of a noninteracting system of the same density as $\Efo={\kfo}^2/2m$.
For simplicity we only consider the limit $w\to 0$ corresponding to a strict 2D system.
A non-zero value of $w$ leads to qualitatively the same physics with only
 a small shift in the critical coupling strength for stripe formation, as long as $\kfo w\ll1$~\cite{Block2012}.

\section{Mean-field theory }
The instability towards forming a striped phase is signalled by a zero frequency pole in the density-density response function
at a given wave number ${\mathbf q}_c$~\cite{Block2012}. Close to the transition to the normal phase, the density modulation
is dominated by the lowest Fourier components ${\mathbf q}_c$ and $-{\mathbf q}_c$, and we can write
\begin{equation}
\rho({\mathbf r})=\rho_0+\rho_1\cos({\mathbf q}_c\cdot{\mathbf r}-u)
\label{eq:stripe}
\end{equation}
with $A^{-1}\sum_{\kv}\mean{\cd{\kv}\co{\kv}}=\rho_0$ and
$A^{-1}\sum_{\kv}\mean{\cd{\kv}\co{\kv+\qcv}}=\rho_1\exp(iu)/2$. Here, $u$ is
the phase shift of the wave and $A$ is the area of the system.
The lowest Fourier components ${\mathbf q}_c$ and $-{\mathbf q}_c$ also dominate deeper into the striped phase, and in the following we therefore neglect the contribution
of higher harmonics to $\rho({\mathbf r})$. Using Wick's theorem to expand the interaction part of the Hamiltonian (\ref{eq:Hamiltonian}),
we construct a mean-field theory by including $\mean{\cd{\kv}\co{\kv'}}$ for $\kv=\kpv$ and $\kv=\kpv\pm\qcv$
which yields the mean-field Hamiltonian
\begin{equation}
\HMF =\sum_{\kv} \epsHF(\kv)\cd{\kv}\co{\kv}+\sum_{\kv} [\ordpar(\kv)\cd{\kv+\qcv}\co{\kv} + \text{h.c.}]. \label{eq:HMF}
\end{equation}
Here,
\begin{equation}
\epsHF(\kv)=\frac{k^2}{2m}+\frac{1}{A}\sum_{\kv'}[V(0)-V({\mathbf k}-{\mathbf k}')]\mean{\cd{\kv'}\co{\kv'}}
\label{eq:epsilon}
\end{equation}
is the Hartree-Fock single particle energy and
\begin{equation}
\ordpar(\kv)=\frac{1}{A}\sum_{\kv'}[V(\qcv)-V({\mathbf k}-{\mathbf k}')]\mean{\cd{\kv'}\co{\kv'+\qcv}}.
\label{eq:orderparameter}
\end{equation}
As usual, these parameters have to be determined self-consistently. This is complicated significantly by the fact that $\ordpar(\kv)$ is a function
 of $\kv$, since it includes the exchange interaction $V({\mathbf k}-{\mathbf k}')$. It is however crucial to include exchange, since
 it is known to lead to important effects such as the collapse of the system, a large deformation of the Fermi surface \cite{Miyakawa2008,Bruun2008}, and a significant
 change in the critical coupling strength for the stripe instability~\cite{Babadi2011a,Sieberer2011,Block2012}. All these effects are recovered
  in our calculation as will be discussed below. For the self-consistent solution, we choose $u=0$ in (\ref{eq:stripe}) which corresponds to $\ordpar(\kv)$ real.

\subsection{Band structure}
In the striped phase, the translational symmetry is spontaneously broken in the direction perpendicular to the stripes in analogy with a classical smectic liquid crystal~\cite{Chaikin1995},
 while it is conserved in the direction along the stripes.
Each dipole experiences the mean-field potential from the other dipoles which is periodic in the direction perpendicular to the stripes.
It follows that a dipole with momentum $\kv$ is coupled only to dipoles with momenta $\kv\pm n\qcv$ with $n$ an integer.
This allows us to think of $k$-space in terms of having a 1D Brillouin zone structure in the direction of $\qcv$ and an unrestricted $k$-space in the direction perpendicular to $\qcv$.
We therefore partition the 2D $k$-space into slices of width $q_c$, by starting with the first Brillouin zone
$B_0$ defined as the points $\kv$ such that $-q_c/2<\kv\cdot \hat{\vekk{q}}_c\le q_c/2$, where $\hat{\vekk{q}}_c$ is the unit vector in the direction of $\qcv$. Then any $k$-space point $\kv$ can be uniquely written as
\begin{equation}
 \label{eq:krep}
\kv=\kpv+n\qcv\quad \text{where $\kpv\in B_0$ and $n\in\mathbb{Z}$.}
\end{equation}
The higher order zones are denoted by $B_n=\{\kv \in \mathbb{R}^2|\exists \kpv \in B_0: \kv=\kpv+n\qcv\}$, and the full $k$-space is the disjoint
union of all $B_n$'s. With this partitioning each $\kv$ state only couples to itself and precisely one state in each of the two neighbouring Brillouin zones,
and the mean-field Hamiltonian \eqref{eq:HMF} can be written as a sum over Hamiltonians for each $\kv$ in the first Brillouin zone: $\HMF=\sum_{\kv\in B_0} \hat{\mathbf c}^\dagger_{\mathbf k} \matr{H}({\kv})\hat{\mathbf c}_{\mathbf k}$. Here $\matr{H}({\kv})$ is a tridiagonal matrix describing the coupling between states with momenta $\kv+n\qcv$ and
$\hat{\mathbf c}^\dagger_{\mathbf k}=(\dots,\cd{\kv-n\qcv},\dots,\cd{\kv},\cd{\kv+\qcv},\dots,\cd{\kv+n\qcv},\dots )$.
So the mean-field Hamiltonian can be diagonalised for each $\kv$ in $B_0$ separately. Note however that the self-consistent averages in
(\ref{eq:epsilon})-(\ref{eq:orderparameter}) are determined by summing over all $\kv$, thereby coupling different $\kv$'s.

For each $\kv$ in the first Brillouin zone, we diagonalise the Hamiltonian $\matr{H}({\kv})$ by introducing the
 quasi-particle operator $\hat{\mathbf \gamma}_{\mathbf k}={\mathbf U}({\mathbf k})^{\dagger}\hat{\mathbf c}_{\mathbf k}$ so that
${\mathbf U}({\mathbf k})^{\dagger}\matr{H}({\kv}){\mathbf U}({\mathbf k})={\mathbf D}({\mathbf k})$ is a diagonal matrix with quasi-particle energies $E_i(\kv)$ on the diagonal.
The thermal average of the new operators for $\kv,\kpv\in B_0$ is given by $\mean{\gd{\kv,i}\go{\kpv,j}}=\delta_{\kv,\kpv}\delta_{i,j}f[E_i(\kv)]$, where
$f(x)=[\exp\beta (x-\mu)+1]^{-1}$ is
 the Fermi-Dirac distribution and $\mu$ is the chemical potential. We can then calculate the Hartree-Fock energies and the order parameter selfconsistently from the diagonal Hamiltonian. The chemical potential is determined by keeping the density of dipoles fixed.

With the geometry illustrated in Fig.~\ref{fig:system}, the stripes are parallel to the $k_x$-axis, since this minimises the repulsion between the dipoles. Thus, the
vector $\qcv$ is parallel to the $k_y$-axis corresponding to $\phi_c=\pi/2$, where $\phi$ is the azimuthal angle between a vector in $\kv$-space and the $k_x$-axis.
We furthermore choose $q_c=2k_F(\phi_c,\theta,g)$ since we expect this to lead to the lowest critical coupling strength for
 stripe formation, as the density wave can then be formed by
particle-hole excitations around the Fermi surface with no cost in kinetic energy.
Here $k_F(\phi,\theta,g)$ is the length of the Fermi vector in the normal phase for interaction strength $g$ and dipole tilting $\theta$.
It depends on the angle $\phi$, since the Fermi surface is deformed  by the dipole-dipole interaction forming an elliptical shape, see Fig.\ \ref{fig:k_space_occupation}. We calculate the deformation using the variational method based on Hartree Fock theory described in \cite{Bruun2008}.

\subsection{Three band theory}\label{sec:three-band-theory}
To proceed, we reduce the numerical complexity by truncating the matrix $\matr{H}({\kv})$ which is to be diagonalized at each point in the first Brillouin zone.
As shown in the appendix~\ref{app:k-space}, the condition for the stripe instability obtained from calculating the density-density response function
in the conserving HFA involves for $T=0$
 momenta only in the three lowest Brillouin zones $B_n$ with $n=-1,0,1$. Thus, we include these three Brillouin zones in our calculations whereas higher energy
zones are neglected. In this way we recover the instability line obtained in Ref.\ \cite{Block2012}.
Higher Brillouin zones contribute in the striped phase or at non-zero temperature, but as long as $T\ll \epsilon_F$ and $\ordpar(\kv)\ll \epsilon_F$ their contribution
is negligible. We shall later demonstrate numerically that including the lowest three zones only is an excellent approximation for the parameters chosen.

The mean-field Hamiltonian for a given $\kv\in B_0$ is then
\begin{equation}
\label{Hmatr}
\matr{H}({\kv})=\begin{bmatrix}
\epsilon_{\kv-\qcv} & \ordpar_{\kv-\qcv}^* & 0\\
\ordpar_{\kv-\qcv} & \epsilon_\kv & \ordpar_{\kv}^{*}\\
0&\ordpar_\kv &\epsilon_{\kv+\qcv}
\end{bmatrix}
\end{equation}
and the three quasi particle energy bands $E_{\kv,1}\le E_{\kv,2}\le E_{\kv,3}$ are the eigenvalues of the matrix $\matr{H}({\kv})$.

The self-consistent equations \eqref{eq:epsilon} for the Hartree Fock energy and \eqref{eq:orderparameter} for the order parameter read in terms of the new single particle
eigenstates
\begin{widetext}

\begin{align}
 \epsilon(\kv)=&\frac{k^2}{2m}+\frac{1}{A}\sum_{\kpv\in B_0}\sum_{n=-1}^{1}
 [V(0)-V(\kv-\kpv-n\qcv)]
 \times \sum_{l=1}^3\norm{U({\kpv})_{n+2,l}}^2f(E_{\kpv,l})\label{eq:epsilon3by3}
\\
\ordpar(\kv)=&\frac{1}{A}\sum_{\kpv\in B_0}\left\{[V(\qcv)-V(\kv-\kpv+\qcv)]\vphantom{\sum_1^1}\right.
\times\sum_{l=1}^3U({\kpv})_{1,l}^{*}U({\kpv})_{2,l}f(E_{\kpv,l})\nonumber\\
&+[V(\qcv)-V(\kv-\kpv)]
\left.\times\sum_{l=1}^3U({\kpv})_{2,l}^{*}U({\kpv}))_{3,l}f(E_{\kpv,l})\right\} \label{eq:ordpar3by3},
\end{align}
while the Fourier components of the density read
\begin{equation}
 \rho_0=\frac{1}{A}\sum_{\kv \in B_0}\sum_{l=1}^3f(E_{\kv,l}),\quad \textup{and} \quad
\rho_1=\frac{2}{A}\sum_{\kv \in B_0}\sum_{l=1}^3[U({\kv})_{1,l}^{*}U({\kv})_{2,l}+U({\kv})_{2,l}^{*}U({\kv})_{3,l}]f(E_{\kv,l}). \label{eq:fourier}
\end{equation}
\end{widetext}
The equations \eqref{Hmatr}-\eqref{eq:fourier} are solved self-consistently by discretizing $B_0$ using a
rectangular grid including states up to $\pm 1.1 k_F(\phi=0,g,\theta)$ in the $k_x$-direction. $B_0$ is defined as being infinite in the $k_x$ direction, but there is no coupling in the direction perpendicular to $\qcv$, so the states with $k_x$ outside the Fermi surface of the normal phase are not occupied. The grid is $n_{k_x}\times n_{k_y}=101 \times 161$ points with an increased density of points near the edges. The iteration procedure is as follows: For each $\kv-$point the $3\times3$ matrix $\matr{U}(\kv)$ and the eigenenergies $E_{i}(\kv)$ are formed by diagonalization of $\matr{H}(\kv)$ computed using the current estimates to $\ordpar$ and $\epsilon$. Then $\mu$ is calculated such that the density is constant and finally the new estimates to $\rho_1$, $\ordpar$, and $\epsilon$ are calculated from \eqref{eq:epsilon3by3}-\eqref{eq:fourier}. The iteration is terminated when the absolute change in $\rho_1$ is less than $10^{-6}$, while the maximum absolute change in any $\kv-$point for $\ordpar$ and $\epsilon$ is less than $10^{-3} \Efo$ and $5\cdot 10^{-3} \Efo$, respectively.

\section{Results}
We now discuss the main results of our numerical calculations yielding self-consistent solutions to \eqref{Hmatr}-\eqref{eq:ordpar3by3}.

\subsection{Stripe amplitude}
\begin{figure}[htb]
\def\svgwidth{.98\columnwidth}
\begingroup%
  \makeatletter%
  \providecommand\color[2][]{%
    \errmessage{(Inkscape) Color is used for the text in Inkscape, but the package 'color.sty' is not loaded}%
    \renewcommand\color[2][]{}%
  }%
  \providecommand\transparent[1]{%
    \errmessage{(Inkscape) Transparency is used (non-zero) for the text in Inkscape, but the package 'transparent.sty' is not loaded}%
    \renewcommand\transparent[1]{}%
  }%
  \providecommand\rotatebox[2]{#2}%
  \ifx\svgwidth\undefined%
    \setlength{\unitlength}{420bp}%
    \ifx\svgscale\undefined%
      \relax%
    \else%
      \setlength{\unitlength}{\unitlength * \real{\svgscale}}%
    \fi%
  \else%
    \setlength{\unitlength}{\svgwidth}%
  \fi%
  \global\let\svgwidth\undefined%
  \global\let\svgscale\undefined%
  \makeatother%
  \begin{picture}(1,0.75)%
    \put(0,0){\includegraphics[width=\unitlength]{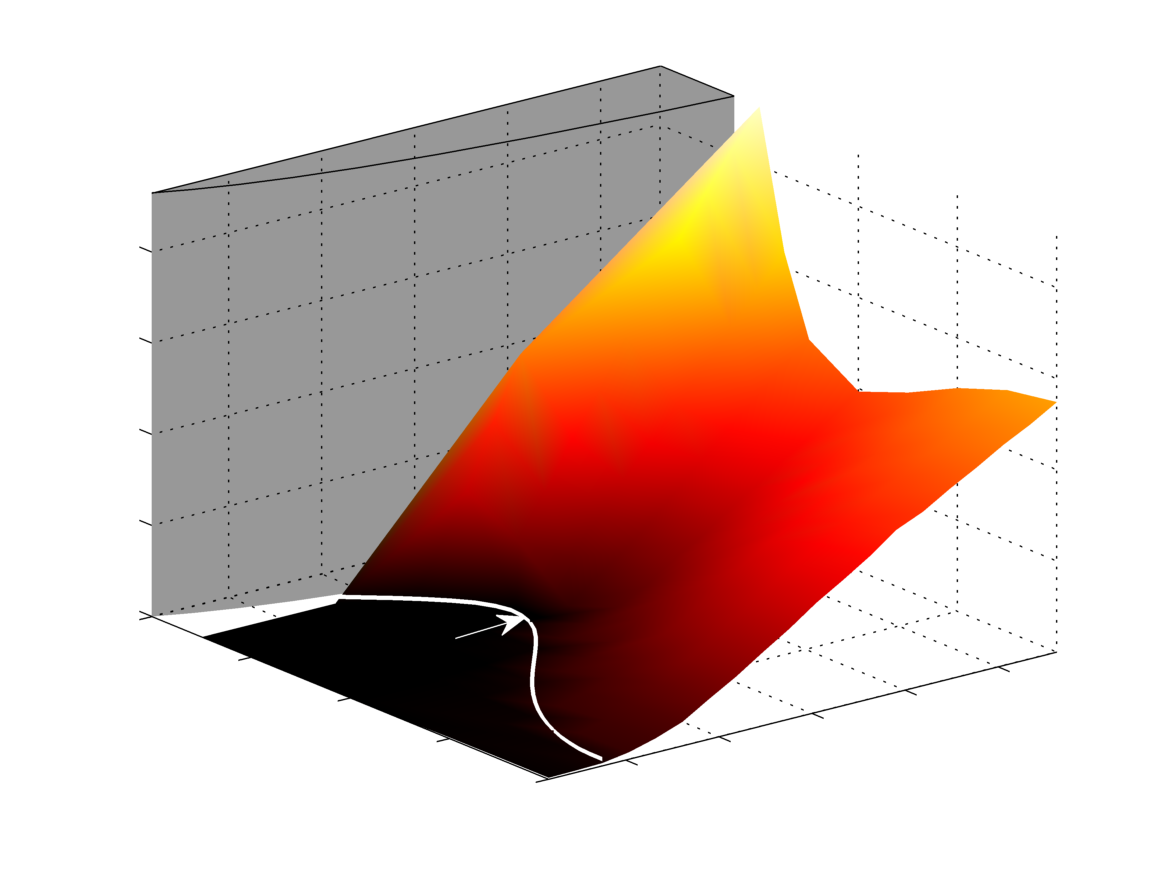}}%
    \put(0.66660762,0.033){\color[rgb]{0,0,0}\makebox(0,0)[lb]{\smash{$g$}}}%
    \put(0.29339238,0.033){\color[rgb]{0,0,0}\makebox(0,0)[rb]{\smash{$\theta/\pi$}}}%
    \put(0.051,0.40152952){\color[rgb]{0,0,0}\rotatebox{90}{\makebox(0,0)[b]{\smash{$\rho_{1}/\rho_{0}$}}}}%
    \put(0.17773524,0.48005714){\color[rgb]{1,1,1}\makebox(0,0)[lb]{\smash{\scriptsize Collapse region}}}%
    \put(0.55287619,0.07097333){\color[rgb]{0,0,0}\makebox(0,0)[b]{\smash{\scriptsize 0.6}}}%
    \put(0.63256571,0.09092952){\color[rgb]{0,0,0}\makebox(0,0)[b]{\smash{\scriptsize 0.7}}}%
    \put(0.71225524,0.11088571){\color[rgb]{0,0,0}\makebox(0,0)[b]{\smash{\scriptsize 0.8}}}%
    \put(0.79194667,0.1308419){\color[rgb]{0,0,0}\makebox(0,0)[b]{\smash{\scriptsize 0.9}}}%
    \put(0.87163619,0.1507981){\color[rgb]{0,0,0}\makebox(0,0)[b]{\smash{\scriptsize 1}}}%
    \put(0.44816762,0.05598095){\color[rgb]{0,0,0}\makebox(0,0)[b]{\smash{\scriptsize 0}}}%
    \put(0.36328762,0.09080381){\color[rgb]{0,0,0}\makebox(0,0)[b]{\smash{\scriptsize 0.1}}}%
    \put(0.27840762,0.12562476){\color[rgb]{0,0,0}\makebox(0,0)[b]{\smash{\scriptsize 0.2}}}%
    \put(0.19352762,0.16044762){\color[rgb]{0,0,0}\makebox(0,0)[b]{\smash{\scriptsize 0.3}}}%
    \put(0.10864762,0.19527048){\color[rgb]{0,0,0}\makebox(0,0)[b]{\smash{\scriptsize 0.4}}}%
    \put(0.11300381,0.22089143){\color[rgb]{0,0,0}\makebox(0,0)[rb]{\smash{\scriptsize 0}}}%
    \put(0.11300381,0.29897524){\color[rgb]{0,0,0}\makebox(0,0)[rb]{\smash{\scriptsize 0.1}}}%
    \put(0.11300381,0.37705714){\color[rgb]{0,0,0}\makebox(0,0)[rb]{\smash{\scriptsize 0.2}}}%
    \put(0.11300381,0.45514095){\color[rgb]{0,0,0}\makebox(0,0)[rb]{\smash{\scriptsize 0.3}}}%
    \put(0.11300381,0.53322286){\color[rgb]{0,0,0}\makebox(0,0)[rb]{\smash{\scriptsize 0.4}}}%
    \put(0.39,0.19039238){\color[rgb]{1,1,1}\makebox(0,0)[rb]{\smash{\scriptsize Critical line}}}%
  \end{picture}%
\endgroup%
 \caption{(colour online) The amplitude $\rho_1/\rho_0$ of the density wave as a function of the coupling strength $g$ and tilt angle $\theta$. The white line is the critical line for stripe formation obtained from linear response~\cite{Block2012}, while the shaded region depicts the collapse region~\cite{Bruun2008,Yamaguchi2010}.}
 \label{fig:density_amplitude3D}
\end{figure}
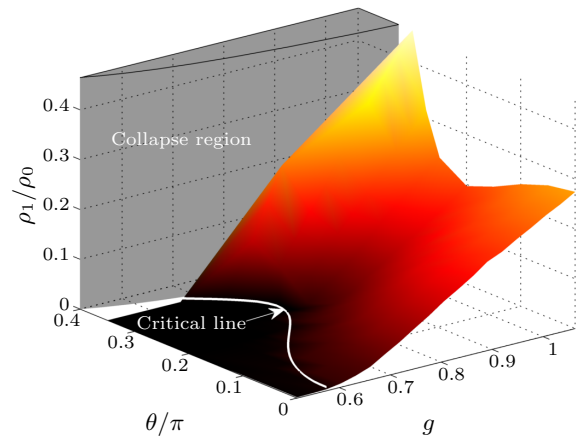

Figure~\ref{fig:density_amplitude3D} shows the amplitude of the density wave $\rho_1/\rho_0$, which we take to be the order parameter of the striped phase,
 as a function of the coupling strength and alignment angle $\theta$.
We clearly see the onset of stripe order beyond a critical coupling strength which depends on the angle. Note that the transition to the broken symmetry phase is
not completely sharp since the discretisation of $k$-space in the numerical calculations
 corresponds to finite size effects, which result in a smooth crossover between the normal and
the striped phase. Taking this small effect into account, the critical coupling strength for the stripe instability obtained here agrees
well with the previous result based on linear response theory~\cite{Block2012}, which is indicated by a white line
in Fig.~\ref{fig:density_amplitude3D}.
This confirms the consistency of our approach and accuracy of the numerics of the present paper. We see that the order parameter
 $\rho_1/\rho_0$ quickly increases with increasing coupling strength resulting in significant density modulations in the striped phase.

For large tilting angles $\theta$ of the dipoles, Hartree-Fock theory predicts that the homogenous state is unstable against density collapse for strong
coupling~\cite{Bruun2008}. This instability is also predicted using a different theoretical approach
obtaining almost the same critical coupling strength for collapse~\cite{Parish2012}. Since a broken symmetry in general leads to a gap in the
spectrum thereby making the system less compressible, an interesting question is
whether the striped phase stabilises the gas against this collapse. However, we do not find any numerical evidence of such a stabilising effect. On the contrary,
our numerical calculations do not converge in the region where a homogeneous phase is predicted to collapse~\cite{Bruun2008,Yamaguchi2010}, which is
 depicted by a shaded region on Fig.~\ref{fig:density_amplitude3D}. This indicates that stripe order does not stabilise the systems against collapse. We speculate that
the reason  is that the systems remains gapless in certain regions of the Fermi surface in the striped phase,
 as we shall discuss in detail below.

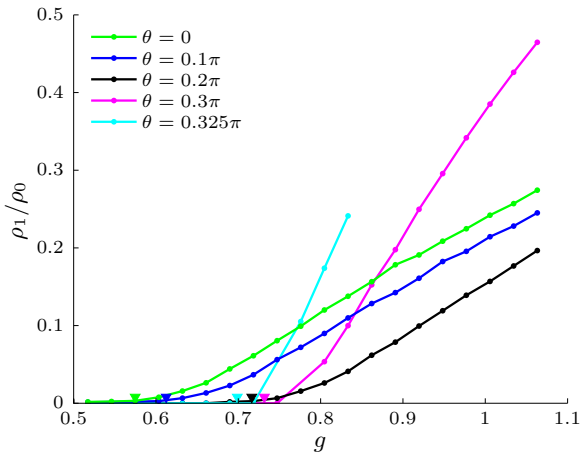
\begin{figure}[htb]
\def\svgwidth{.98\columnwidth}
\begingroup%
  \makeatletter%
  \providecommand\color[2][]{%
    \errmessage{(Inkscape) Color is used for the text in Inkscape, but the package 'color.sty' is not loaded}%
    \renewcommand\color[2][]{}%
  }%
  \providecommand\transparent[1]{%
    \errmessage{(Inkscape) Transparency is used (non-zero) for the text in Inkscape, but the package 'transparent.sty' is not loaded}%
    \renewcommand\transparent[1]{}%
  }%
  \providecommand\rotatebox[2]{#2}%
  \ifx\svgwidth\undefined%
    \setlength{\unitlength}{420bp}%
    \ifx\svgscale\undefined%
      \relax%
    \else%
      \setlength{\unitlength}{\unitlength * \real{\svgscale}}%
    \fi%
  \else%
    \setlength{\unitlength}{\svgwidth}%
  \fi%
  \global\let\svgwidth\undefined%
  \global\let\svgscale\undefined%
  \makeatother%
  \begin{picture}(1,0.75)%
    \put(0,0){\includegraphics[width=\unitlength]{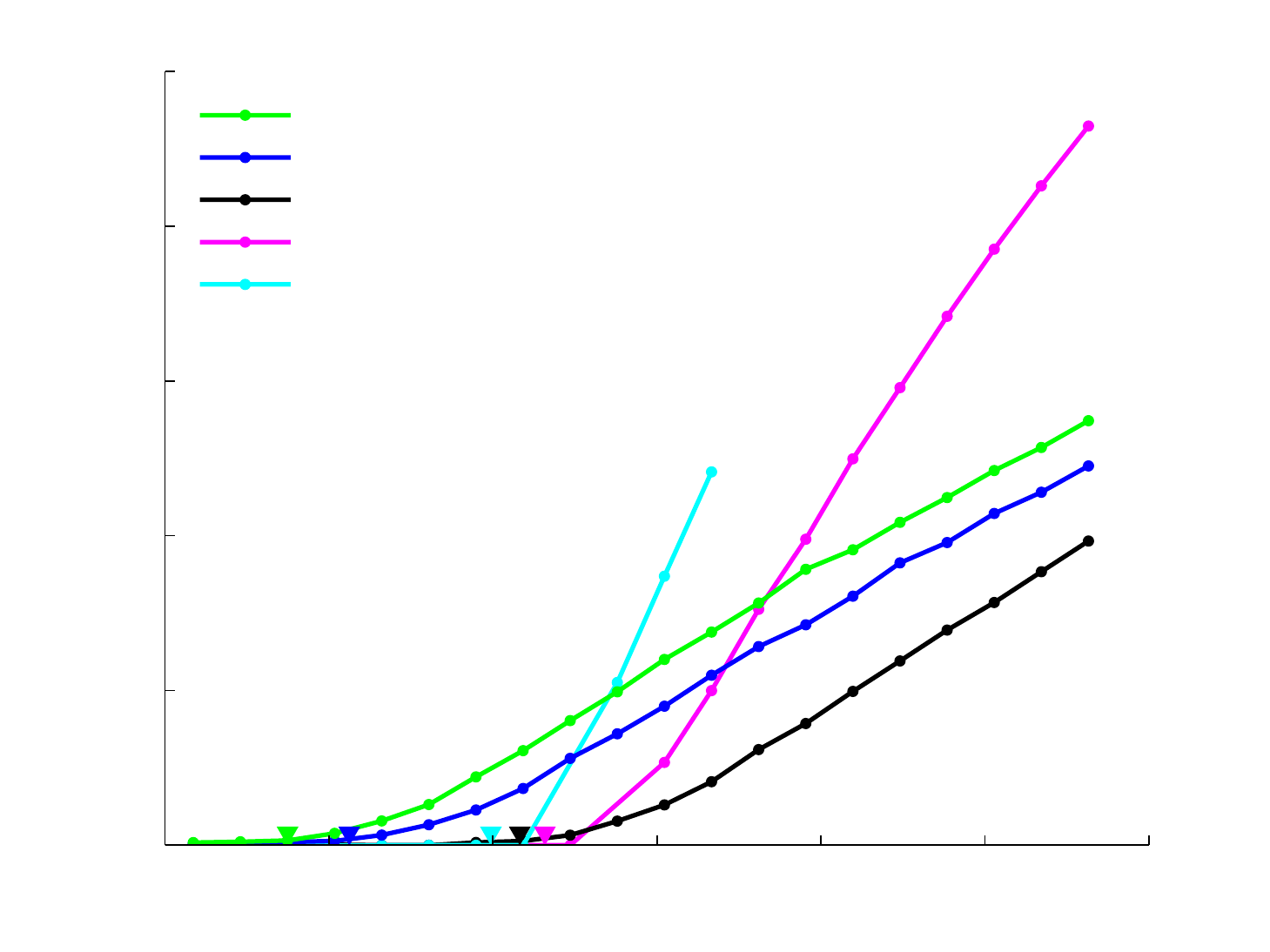}}%
    \put(0.51660762,0.01476){\color[rgb]{0,0,0}\makebox(0,0)[b]{\smash{$g$}}}%
    \put(0.05,0.38705143){\color[rgb]{0,0,0}\rotatebox{90}{\makebox(0,0)[b]{\smash{$\rho_1/\rho_0$}}}}%
    \put(0.13,0.05066095){\color[rgb]{0,0,0}\makebox(0,0)[b]{\smash{\scriptsize 0.5}}}%
    \put(0.25916571,0.05066095){\color[rgb]{0,0,0}\makebox(0,0)[b]{\smash{\scriptsize 0.6}}}%
    \put(0.38833333,0.05066095){\color[rgb]{0,0,0}\makebox(0,0)[b]{\smash{\scriptsize 0.7}}}%
    \put(0.51750095,0.05066095){\color[rgb]{0,0,0}\makebox(0,0)[b]{\smash{\scriptsize 0.8}}}%
    \put(0.64666667,0.05066095){\color[rgb]{0,0,0}\makebox(0,0)[b]{\smash{\scriptsize 0.9}}}%
    \put(0.77583238,0.05066095){\color[rgb]{0,0,0}\makebox(0,0)[b]{\smash{\scriptsize 1}}}%
    \put(0.905,0.05066095){\color[rgb]{0,0,0}\makebox(0,0)[b]{\smash{\scriptsize 1.1}}}%
    \put(0.11449905,0.07589333){\color[rgb]{0,0,0}\makebox(0,0)[rb]{\smash{\scriptsize 0}}}%
    \put(0.11450095,0.19785714){\color[rgb]{0,0,0}\makebox(0,0)[rb]{\smash{\scriptsize 0.1}}}%
    \put(0.11449905,0.31982095){\color[rgb]{0,0,0}\makebox(0,0)[rb]{\smash{\scriptsize 0.2}}}%
    \put(0.11449905,0.44178476){\color[rgb]{0,0,0}\makebox(0,0)[rb]{\smash{\scriptsize 0.3}}}%
    \put(0.11449905,0.56375048){\color[rgb]{0,0,0}\makebox(0,0)[rb]{\smash{\scriptsize 0.4}}}%
    \put(0.11449905,0.68571429){\color[rgb]{0,0,0}\makebox(0,0)[rb]{\smash{\scriptsize 0.5}}}%
    \put(0.23690476,0.64970286){\color[rgb]{0,0,0}\makebox(0,0)[lb]{\smash{\scriptsize$\theta=0$}}}%
    \put(0.23690476,0.61577333){\color[rgb]{0,0,0}\makebox(0,0)[lb]{\smash{\scriptsize$\theta=0.1\pi$}}}%
    \put(0.23690476,0.58184571){\color[rgb]{0,0,0}\makebox(0,0)[lb]{\smash{\scriptsize$\theta=0.2\pi$}}}%
    \put(0.23690476,0.54970286){\color[rgb]{0,0,0}\makebox(0,0)[lb]{\smash{\scriptsize$\theta=0.3\pi$}}}%
    \put(0.23690476,0.51577333){\color[rgb]{0,0,0}\makebox(0,0)[lb]{\smash{\scriptsize$\theta=0.325\pi$}}}%
  \end{picture}%
\endgroup%
 \caption{(colour online) $\rho_1/\rho_0$ as a function of $g$ for various tilt angles $\theta$. 
 The critical value for stripe formation~\cite{Block2012} for each angle is marked by a triangle $\triangledown$ of the corresponding color.}
 \label{fig:density_amplitude2D}
\end{figure}

In Fig.~\ref{fig:density_amplitude2D}, we plot the stripe order parameter $\rho_1/\rho_0$ as a function of $g$ for various tilt angles $\theta$.
These curves correspond to cuts along constant $\theta$ in Fig.~\ref{fig:density_amplitude3D}. They clearly
illustrate that apart from finite size effects, the critical coupling strength for the
onset of pairing agrees well with what is obtained from a linear response~\cite{Block2012}. An interesting effect is that
 the stripe amplitude increases faster for larger angles $\theta$, where the interaction is increasingly anisotropic and the system approaches the collapse instability.

\subsection{Momentum distribution}
\begin{figure}[htb]
\def\svgwidth{.98\columnwidth} 
\begingroup%
  \makeatletter%
  \providecommand\color[2][]{%
    \errmessage{(Inkscape) Color is used for the text in Inkscape, but the package 'color.sty' is not loaded}%
    \renewcommand\color[2][]{}%
  }%
  \providecommand\transparent[1]{%
    \errmessage{(Inkscape) Transparency is used (non-zero) for the text in Inkscape, but the package 'transparent.sty' is not loaded}%
    \renewcommand\transparent[1]{}%
  }%
  \providecommand\rotatebox[2]{#2}%
  \ifx\svgwidth\undefined%
    \setlength{\unitlength}{440.2504bp}%
    \ifx\svgscale\undefined%
      \relax%
    \else%
      \setlength{\unitlength}{\unitlength * \real{\svgscale}}%
    \fi%
  \else%
    \setlength{\unitlength}{\svgwidth}%
  \fi%
  \global\let\svgwidth\undefined%
  \global\let\svgscale\undefined%
  \makeatother%
  \begin{picture}(1,0.75809221)%
    \put(0,0){\includegraphics[width=\unitlength]{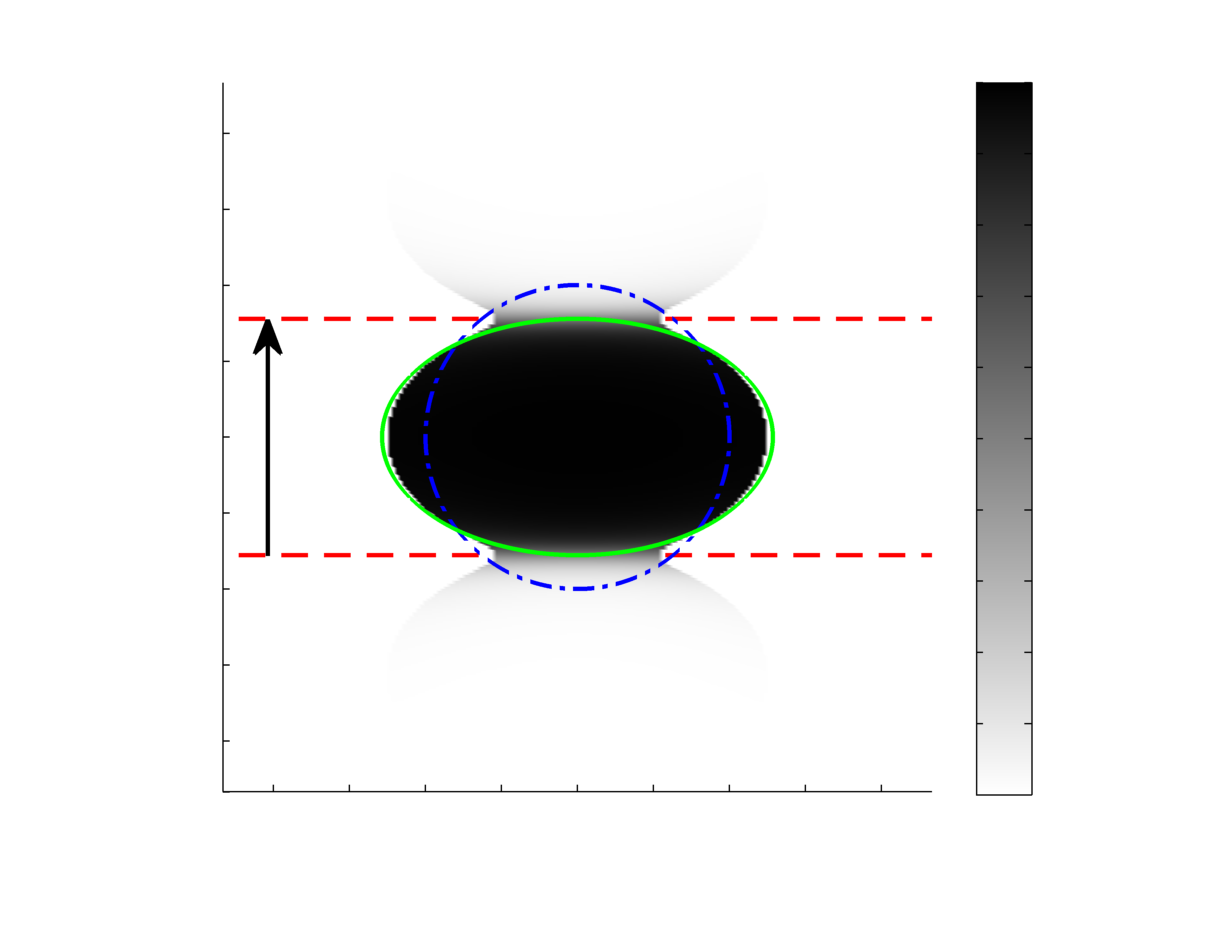}}%
    \put(0.47126452,0.04474546){\color[rgb]{0,0,0}\makebox(0,0)[b]{\smash{$k_x/\kfo$}}}%
    \put(0.105,0.39991424){\color[rgb]{0,0,0}\rotatebox{90}{\makebox(0,0)[b]{\smash{$k_y/\kfo$}}}}
    \put(0.69447749,0.58185797){\color[rgb]{1,0,0}\makebox(0,0)[lb]{\smash{$B_1$}}}%
    \put(0.69447749,0.38850074){\color[rgb]{1,0,0}\makebox(0,0)[lb]{\smash{$B_0$}}}%
    \put(0.69447749,0.19514532){\color[rgb]{1,0,0}\makebox(0,0)[lb]{\smash{$B_{-1}$}}}%
    \put(0.23042341,0.38850074){\color[rgb]{0,0,0}\makebox(0,0)[lb]{\smash{$\mathbf{q}_c$}}}%
    \put(0.22353278,0.07902959){\color[rgb]{0,0,0}\makebox(0,0)[b]{\smash{\scriptsize -2}}}%
    \put(0.28567923,0.07902959){\color[rgb]{0,0,0}\makebox(0,0)[b]{\smash{\scriptsize -1.5}}}%
    \put(0.34782568,0.07902959){\color[rgb]{0,0,0}\makebox(0,0)[b]{\smash{\scriptsize -1}}}%
    \put(0.40997213,0.07902959){\color[rgb]{0,0,0}\makebox(0,0)[b]{\smash{\scriptsize -0.5}}}%
    \put(0.47211676,0.07902959){\color[rgb]{0,0,0}\makebox(0,0)[b]{\smash{\scriptsize 0}}}%
    \put(0.53426321,0.07902959){\color[rgb]{0,0,0}\makebox(0,0)[b]{\smash{\scriptsize 0.5}}}%
    \put(0.59640966,0.07902959){\color[rgb]{0,0,0}\makebox(0,0)[b]{\smash{\scriptsize 1}}}%
    \put(0.65855611,0.07902959){\color[rgb]{0,0,0}\makebox(0,0)[b]{\smash{\scriptsize 1.5}}}%
    \put(0.72070256,0.07902959){\color[rgb]{0,0,0}\makebox(0,0)[b]{\smash{\scriptsize 2}}}%
    \put(0.17048297,0.14451596){\color[rgb]{0,0,0}\makebox(0,0)[rb]{\smash{\scriptsize -2}}}%
    \put(0.17048297,0.20666241){\color[rgb]{0,0,0}\makebox(0,0)[rb]{\smash{\scriptsize -1.5}}}%
    \put(0.17048297,0.26880886){\color[rgb]{0,0,0}\makebox(0,0)[rb]{\smash{\scriptsize -1}}}%
    \put(0.17048297,0.3309553){\color[rgb]{0,0,0}\makebox(0,0)[rb]{\smash{\scriptsize -0.5}}}%
    \put(0.17048297,0.39310175){\color[rgb]{0,0,0}\makebox(0,0)[rb]{\smash{\scriptsize 0}}}%
    \put(0.17048297,0.4552482){\color[rgb]{0,0,0}\makebox(0,0)[rb]{\smash{\scriptsize 0.5}}}%
    \put(0.17048297,0.51739465){\color[rgb]{0,0,0}\makebox(0,0)[rb]{\smash{\scriptsize 1}}}%
    \put(0.17048297,0.5795411){\color[rgb]{0,0,0}\makebox(0,0)[rb]{\smash{\scriptsize 1.5}}}%
    \put(0.17048297,0.64168755){\color[rgb]{0,0,0}\makebox(0,0)[rb]{\smash{\scriptsize 2}}}%
    \put(0.8461657,0.09710476){\color[rgb]{0,0,0}\makebox(0,0)[lb]{\smash{\scriptsize 0}}}%
    \put(0.8461657,0.15570777){\color[rgb]{0,0,0}\makebox(0,0)[lb]{\smash{\scriptsize 0.1}}}%
    \put(0.8461657,0.21431078){\color[rgb]{0,0,0}\makebox(0,0)[lb]{\smash{\scriptsize 0.2}}}%
    \put(0.8461657,0.27291379){\color[rgb]{0,0,0}\makebox(0,0)[lb]{\smash{\scriptsize 0.3}}}%
    \put(0.8461657,0.3315168){\color[rgb]{0,0,0}\makebox(0,0)[lb]{\smash{\scriptsize 0.4}}}%
    \put(0.8461657,0.39011981){\color[rgb]{0,0,0}\makebox(0,0)[lb]{\smash{\scriptsize 0.5}}}%
    \put(0.8461657,0.44872283){\color[rgb]{0,0,0}\makebox(0,0)[lb]{\smash{\scriptsize 0.6}}}%
    \put(0.8461657,0.50732584){\color[rgb]{0,0,0}\makebox(0,0)[lb]{\smash{\scriptsize 0.7}}}%
    \put(0.8461657,0.56592885){\color[rgb]{0,0,0}\makebox(0,0)[lb]{\smash{\scriptsize 0.8}}}%
    \put(0.8461657,0.62453186){\color[rgb]{0,0,0}\makebox(0,0)[lb]{\smash{\scriptsize 0.9}}}%
    \put(0.8461657,0.68313487){\color[rgb]{0,0,0}\makebox(0,0)[lb]{\smash{\scriptsize 1}}}%
  \end{picture}%
\endgroup%
 \caption{(colour online) The momentum distribution $\langle\cd{\kv}\co{\kv}\rangle$ in the striped phase for $g=1.01$, $\theta=0.3 \pi$ where $\rho_1/\rho_0=0.385$, plotted in the three first Brillouin zones $B_{-1},B_0$, and $B_{1}$. 
 The elliptical shape of the underlying Fermi sea (solid green) of the homogenous phase and the circular Fermi sea (dashed blue) for a non-interacting system are also shown. }
 \label{fig:k_space_occupation}
\end{figure}
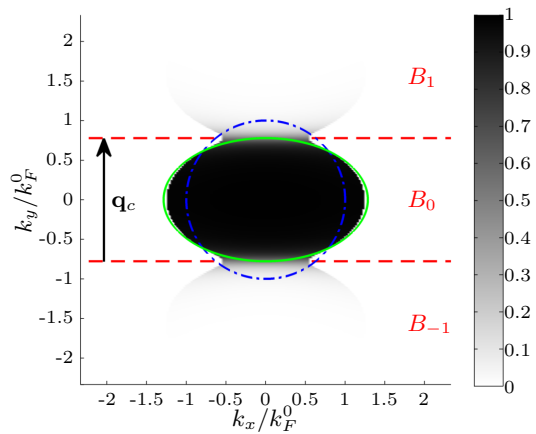

The momentum distribution $\langle\cd{\kv}\co{\kv}\rangle$ of the system can be measured in a time-of-flight (TOF) experiment, and we now analyse how
this can be used to detect the striped phase. In Fig.\ \ref{fig:k_space_occupation}, we plot $\langle\cd{\kv}\co{\kv}\rangle$ for $g=1.01$ and $\theta=0.3\pi$ which
corresponds to a fairly large stripe amplitude $\rho_1/\rho_0=0.385$. First, we notice that the momentum distribution is strongly anisotropic
 in agreement with what is found for the normal phase~\cite{Miyakawa2008,Bruun2008}.
 We plot in Fig.\ \ref{fig:k_space_occupation} an elliptical approximation for the Fermi sea as
 calculated from a variational Hartree Fock theory for the normal phase, as described in \cite{Bruun2008}. We see that the Fermi sea for the striped phase
 has almost the same underlying elliptical shape. To illustrate the significant Fermi surface deformation,
 we also plot the circular Fermi sea of a noninteracting system with the same density.
 In addition to the elliptical shape of the Fermi sea, the striped phase is characterised by a smearing out of the momentum distribution in the regions
 located around the edge of the Fermi sea with $\phi\simeq \pm \pi/2$, i.e.\ $\kv\simeq \pm \qcv/2$. This is because the states with momenta $\kv$ and
$\kv\mp \qcv$ are nearly degenerate in these regions where the Fermi surface is near the edge of the first Brillouin zones.
The resulting strong mixing of the momentum states means that the quasiparticles do not have a well-defined momentum.
Note that these regions are enlarged due to the underlying elliptical shape of the Fermi 
sea creating a "nesting" effect in analogy with lattice systems. It follows from this nesting that 
 the stripe order is enhanced by the elliptical shape of the Fermi sea. 
In total, Fig.\ \ref{fig:k_space_occupation} clearly demonstrates that 
the striped phase can be detected in a TOF experiment by the characteristic shape of its Fermi sea. 

Finally, Fig.\ \ref{fig:k_space_occupation} shows that the population in the Brillouin zones $B_n$ with $n=\pm 1$ is very small. This 
 confirms that the three band approximation is accurate in the striped phase for the parameters used. 

\subsection{Quasi-particle energies}
\begin{figure}[htb]
\def\svgwidth{.98\columnwidth}
\begingroup%
  \makeatletter%
  \providecommand\color[2][]{%
    \errmessage{(Inkscape) Color is used for the text in Inkscape, but the package 'color.sty' is not loaded}%
    \renewcommand\color[2][]{}%
  }%
  \providecommand\transparent[1]{%
    \errmessage{(Inkscape) Transparency is used (non-zero) for the text in Inkscape, but the package 'transparent.sty' is not loaded}%
    \renewcommand\transparent[1]{}%
  }%
  \providecommand\rotatebox[2]{#2}%
  \ifx\svgwidth\undefined%
    \setlength{\unitlength}{311.2504bp}%
    \ifx\svgscale\undefined%
      \relax%
    \else%
      \setlength{\unitlength}{\unitlength * \real{\svgscale}}%
    \fi%
  \else%
    \setlength{\unitlength}{\svgwidth}%
  \fi%
  \global\let\svgwidth\undefined%
  \global\let\svgscale\undefined%
  \makeatother%
  \begin{picture}(1,0.87469896)%
    \put(0,0){\includegraphics[width=\unitlength]{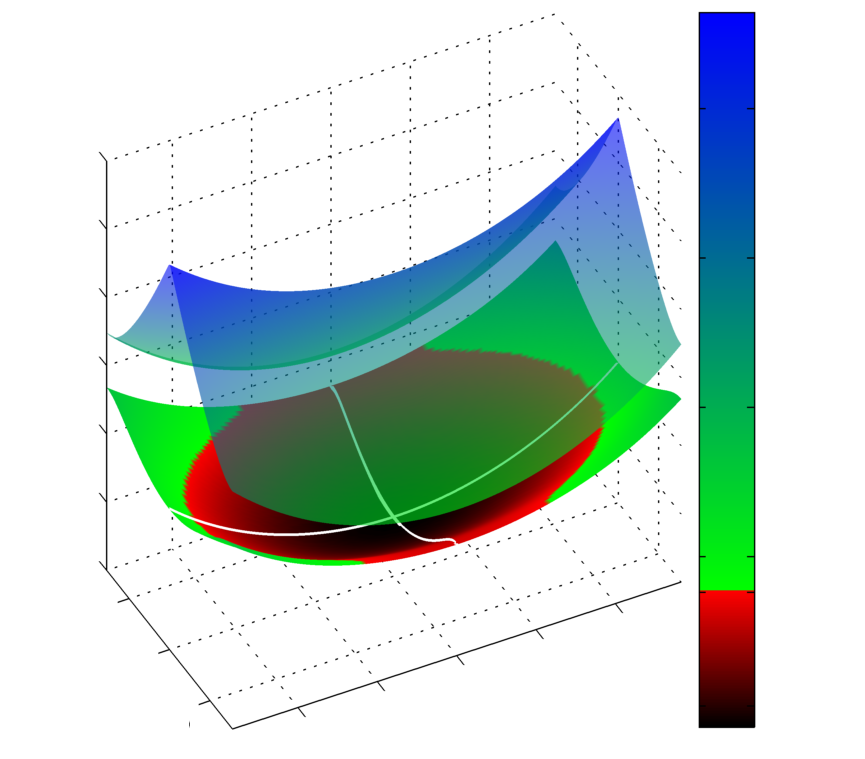}}%
    \put(0.53690275,0.01370989){\color[rgb]{0,0,0}\makebox(0,0)[lb]{\smash{$k_x/\kfo$}}}%
    \put(0.14859804,0.0525108){\color[rgb]{0,0,0}\makebox(0,0)[rb]{\smash{$k_y/\kfo$}}}%
    \put(0.05655125,0.44976265){\color[rgb]{0,0,0}\rotatebox{90}{\makebox(0,0)[b]{\smash{$E/\Efo$}}}}%
    \put(0.35242877,0.02098378){\color[rgb]{0,0,0}\makebox(0,0)[b]{\smash{\scriptsize -1}}}%
    \put(0.4439615,0.0509635){\color[rgb]{0,0,0}\makebox(0,0)[b]{\smash{\scriptsize -0.5}}}%
    \put(0.5354968,0.08094065){\color[rgb]{0,0,0}\makebox(0,0)[b]{\smash{\scriptsize 0}}}%
    \put(0.62702953,0.11092037){\color[rgb]{0,0,0}\makebox(0,0)[b]{\smash{\scriptsize 0.5}}}%
    \put(0.71856226,0.14090009){\color[rgb]{0,0,0}\makebox(0,0)[b]{\smash{\scriptsize 1}}}%
    \put(0.21754317,0.02951453){\color[rgb]{0,0,0}\makebox(0,0)[b]{\smash{\scriptsize -0.5}}}%
    \put(0.17090291,0.08835333){\color[rgb]{0,0,0}\makebox(0,0)[b]{\smash{\scriptsize 0}}}%
    \put(0.12426522,0.14718955){\color[rgb]{0,0,0}\makebox(0,0)[b]{\smash{\scriptsize 0.5}}}%
    \put(0.11698619,0.215104){\color[rgb]{0,0,0}\makebox(0,0)[rb]{\smash{\scriptsize -1}}}%
    \put(0.11698619,0.29380077){\color[rgb]{0,0,0}\makebox(0,0)[rb]{\smash{\scriptsize 0}}}%
    \put(0.11698619,0.37249496){\color[rgb]{0,0,0}\makebox(0,0)[rb]{\smash{\scriptsize 1}}}%
    \put(0.11698619,0.45119172){\color[rgb]{0,0,0}\makebox(0,0)[rb]{\smash{\scriptsize 2}}}%
    \put(0.11698619,0.52988849){\color[rgb]{0,0,0}\makebox(0,0)[rb]{\smash{\scriptsize 3}}}%
    \put(0.11698619,0.60858525){\color[rgb]{0,0,0}\makebox(0,0)[rb]{\smash{\scriptsize 4}}}%
    \put(0.11698619,0.68727944){\color[rgb]{0,0,0}\makebox(0,0)[rb]{\smash{\scriptsize 5}}}%
    \put(0.87367853,0.04967065){\color[rgb]{0,0,0}\makebox(0,0)[lb]{\smash{\scriptsize 0}}}%
    \put(0.87367853,0.18087304){\color[rgb]{0,0,0}\makebox(0,0)[lb]{\smash{\scriptsize $\mu$}}}%
    \put(0.87367853,0.22195636){\color[rgb]{0,0,0}\makebox(0,0)[lb]{\smash{\scriptsize $1 \Efo$}}}%
    \put(0.87367853,0.39424465){\color[rgb]{0,0,0}\makebox(0,0)[lb]{\smash{\scriptsize $2 \Efo$}}}%
    \put(0.87367853,0.56653037){\color[rgb]{0,0,0}\makebox(0,0)[lb]{\smash{\scriptsize $3 \Efo$}}}%
    \put(0.87367853,0.73881865){\color[rgb]{0,0,0}\makebox(0,0)[lb]{\smash{\scriptsize $4 \Efo$}}}%
  \end{picture}%
\endgroup%
\caption{(colour online) The two lowest energy bands for $g=1.01$ and $\theta=0.3\pi$. 
Energies below the chemical potential are colored red, energies above the chemical potential are colored green going to blue. The white lines indicate the cuts along $k_x=0$ and $k_y=0$ which are shown in Fig.~\ref{fig:QP_energiesCuts}.}
 \label{fig:QP_energies}
\end{figure}
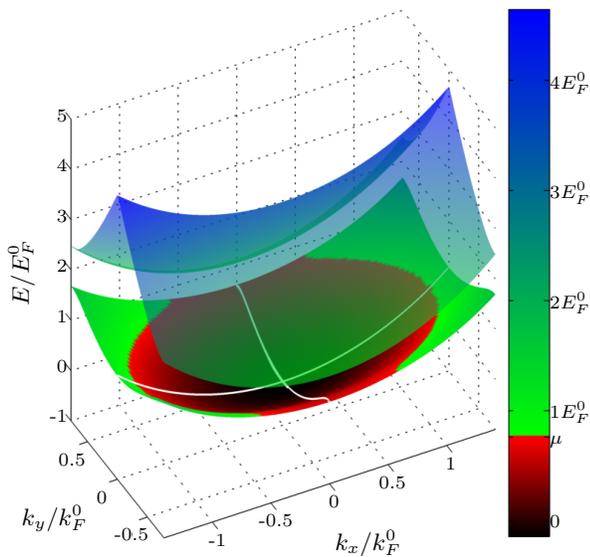

\begin{figure}[htb]
\begin{center}
\def\svgwidth{0.9\columnwidth}
\begingroup%
  \makeatletter%
  \providecommand\color[2][]{%
    \errmessage{(Inkscape) Color is used for the text in Inkscape, but the package 'color.sty' is not loaded}%
    \renewcommand\color[2][]{}%
  }%
  \providecommand\transparent[1]{%
    \errmessage{(Inkscape) Transparency is used (non-zero) for the text in Inkscape, but the package 'transparent.sty' is not loaded}%
    \renewcommand\transparent[1]{}%
  }%
  \providecommand\rotatebox[2]{#2}%
  \ifx\svgwidth\undefined%
    \setlength{\unitlength}{414bp}%
    \ifx\svgscale\undefined%
      \relax%
    \else%
      \setlength{\unitlength}{\unitlength * \real{\svgscale}}%
    \fi%
  \else%
    \setlength{\unitlength}{\svgwidth}%
  \fi%
  \global\let\svgwidth\undefined%
  \global\let\svgscale\undefined%
  \makeatother%
  \begin{picture}(1,0.39492754)%
    \put(0,0){\includegraphics[width=\unitlength]{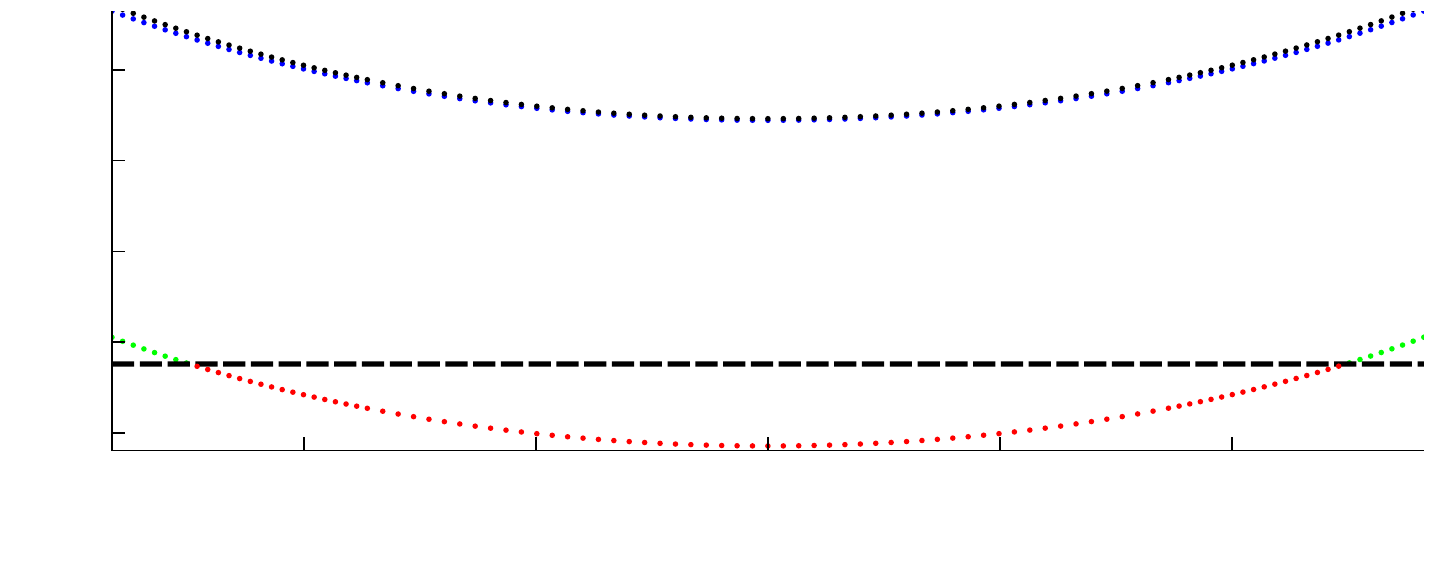}}%
    \put(0.53328889,0.01999034){\color[rgb]{0,0,0}\makebox(0,0)[b]{\smash{$k_x/\kfo$}}}%
    \put(0.02168309,0.23075362){\color[rgb]{0,0,0}\rotatebox{90}{\makebox(0,0)[b]{\smash{$E/\Efo$}}}}%
    \put(0.21154203,0.05221836){\color[rgb]{0,0,0}\makebox(0,0)[b]{\smash{\scriptsize -1}}}%
    \put(0.37286763,0.05221836){\color[rgb]{0,0,0}\makebox(0,0)[b]{\smash{\scriptsize -0.5}}}%
    \put(0.53419324,0.05221836){\color[rgb]{0,0,0}\makebox(0,0)[b]{\smash{\scriptsize 0}}}%
    \put(0.69552077,0.05221836){\color[rgb]{0,0,0}\makebox(0,0)[b]{\smash{\scriptsize 0.5}}}%
    \put(0.85684638,0.05221836){\color[rgb]{0,0,0}\makebox(0,0)[b]{\smash{\scriptsize 1}}}%
    \put(0.05964638,0.08365604){\color[rgb]{0,0,0}\makebox(0,0)[rb]{\smash{\scriptsize 0}}}%
    \put(0.05964638,0.14685217){\color[rgb]{0,0,0}\makebox(0,0)[rb]{\smash{\scriptsize 1}}}%
    \put(0.05964638,0.21004638){\color[rgb]{0,0,0}\makebox(0,0)[rb]{\smash{\scriptsize 2}}}%
    \put(0.05964638,0.27324251){\color[rgb]{0,0,0}\makebox(0,0)[rb]{\smash{\scriptsize 3}}}%
    \put(0.05964638,0.33643865){\color[rgb]{0,0,0}\makebox(0,0)[rb]{\smash{\scriptsize 4}}}%
  \end{picture}%
\endgroup%
\def\svgwidth{0.9\columnwidth}
\begingroup%
  \makeatletter%
  \providecommand\color[2][]{%
    \errmessage{(Inkscape) Color is used for the text in Inkscape, but the package 'color.sty' is not loaded}%
    \renewcommand\color[2][]{}%
  }%
  \providecommand\transparent[1]{%
    \errmessage{(Inkscape) Transparency is used (non-zero) for the text in Inkscape, but the package 'transparent.sty' is not loaded}%
    \renewcommand\transparent[1]{}%
  }%
  \providecommand\rotatebox[2]{#2}%
  \ifx\svgwidth\undefined%
    \setlength{\unitlength}{414bp}%
    \ifx\svgscale\undefined%
      \relax%
    \else%
      \setlength{\unitlength}{\unitlength * \real{\svgscale}}%
    \fi%
  \else%
    \setlength{\unitlength}{\svgwidth}%
  \fi%
  \global\let\svgwidth\undefined%
  \global\let\svgscale\undefined%
  \makeatother%
  \begin{picture}(1,0.39492754)%
    \put(0,0){\includegraphics[width=\unitlength]{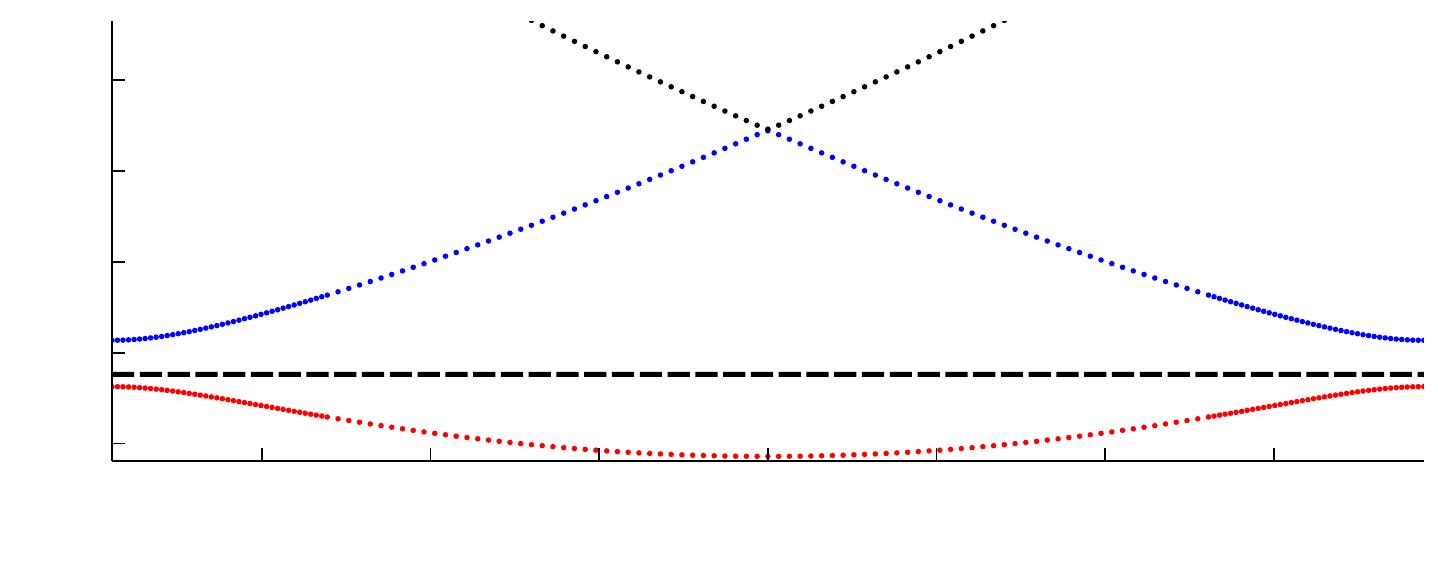}}%
    \put(0.53328889,0.01999034){\color[rgb]{0,0,0}\makebox(0,0)[b]{\smash{$k_y/\kfo$}}}%
    \put(0.02168309,0.23075362){\color[rgb]{0,0,0}\rotatebox{90}{\makebox(0,0)[b]{\smash{$E/\Efo$}}}}%
    \put(0.18221643,0.05221836){\color[rgb]{0,0,0}\makebox(0,0)[b]{\smash{\scriptsize -0.6}}}%
    \put(0.29954203,0.05221836){\color[rgb]{0,0,0}\makebox(0,0)[b]{\smash{\scriptsize -0.4}}}%
    \put(0.41686763,0.05221836){\color[rgb]{0,0,0}\makebox(0,0)[b]{\smash{\scriptsize -0.2}}}%
    \put(0.53419324,0.05221836){\color[rgb]{0,0,0}\makebox(0,0)[b]{\smash{\scriptsize 0}}}%
    \put(0.65152077,0.05221836){\color[rgb]{0,0,0}\makebox(0,0)[b]{\smash{\scriptsize 0.2}}}%
    \put(0.76884638,0.05221836){\color[rgb]{0,0,0}\makebox(0,0)[b]{\smash{\scriptsize 0.4}}}%
    \put(0.88617198,0.05221836){\color[rgb]{0,0,0}\makebox(0,0)[b]{\smash{\scriptsize 0.6}}}%
    \put(0.05964638,0.08365604){\color[rgb]{0,0,0}\makebox(0,0)[rb]{\smash{\scriptsize 0}}}%
    \put(0.05964638,0.14685217){\color[rgb]{0,0,0}\makebox(0,0)[rb]{\smash{\scriptsize 1}}}%
    \put(0.05964638,0.21004638){\color[rgb]{0,0,0}\makebox(0,0)[rb]{\smash{\scriptsize 2}}}%
    \put(0.05964638,0.27324251){\color[rgb]{0,0,0}\makebox(0,0)[rb]{\smash{\scriptsize 3}}}%
    \put(0.05964638,0.33643865){\color[rgb]{0,0,0}\makebox(0,0)[rb]{\smash{\scriptsize 4}}}%
  \end{picture}%
\endgroup%
\end{center}
\caption{(colour online) The two lowest energy bands and the lower part of the third band along $k_y$=0 (top) and $k_x$=0 (bottom) in the first Brillouin zone. Occupied states in the lowest band are indicated by red dots, the chemical potential is indicated by a dashed line, while unoccupied states in the first, second, and third  bands are indicated by green, blue and black dots respectively.}
 \label{fig:QP_energiesCuts}
\end{figure}
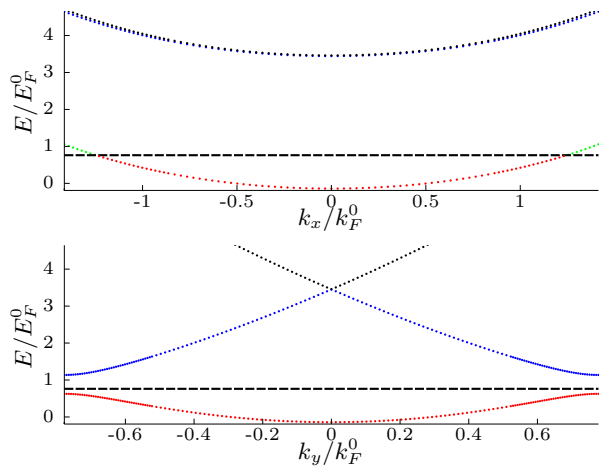

As we discussed, stripe order mixes states with momenta differing by $\qcv$ giving rise to large effects in the regions around $\kv\simeq\pm\qcv/2$.
To examine this effect further, we plot in Fig.\ \ref{fig:QP_energies} the quasiparticle energies for the lowest two bands obtained from 
diagonalising the matrix $\matr{H}({\kv})$ for $g=1.01$, $\theta=0.3\pi$ giving $\rho_1/\rho_0=0.385$.
As expected, we see that the stripe order gives rise to a gap opening up at the Fermi surface in the regions around 
$\kv\simeq\pm\qcv/2$. The system however remains gapless in the other regions of the Fermi surface where the quasiparticle 
energies are perturbed only slightly from their normal phase values. This is illustrated further in Fig.\ \ref{fig:QP_energiesCuts} where 
we plot the quasiparticle energies along cuts defined by $k_x=0$ and $k_y=0$. One clearly sees the gap at the Fermi surface 
for $k_x=0$ whereas there is no gap for $k_y=0$. This explains why the system remains compressible and the stripe order does not 
stabilise the system significantly against collapse. It furthermore opens up the intriguing possibility of forming stripe and superfluid order simultaneously: 
While Cooper pairing is suppressed in the gapped regions around $\kv\simeq \pm\qcv/2$, particles around the gapless regions Fermi surface can still form 
Cooper pairs. Such a phase with both superfluid and density order is a supersolid, and its experimental realisation 
would be a major result, since it has not been unambiguously observed despite 
decades of intense research\cite{Andreev1969,Chester1970,Leggett1970,Kim2004,Chan2008,Capogrosso-Sansone2010}.
It also demonstratres that it is very promising to use dipolar gases to investigate the interplay between quantum liquid crystal phases such a stripes, and superfluid 
pairing, which is a central topic in the physics of strongly correlated systems including cuprate and pnictide 
superconductors~\cite{Kivelson1998,Fradkin2010}. 

\subsection{Momentum correlations}
TOF experiments can also be used to measure correlation functions in quantum gases. Indeed, 
 pair correlations \cite{Greiner2005}, bosonic bunching \cite{Folling2005},
fermionic anti-bunching \cite{Rom2006}, and the Mott-superfluid \cite{Spielman2007} has been measured with 
this technique. We now demonstrate how TOF experiments can be used to detect the formation of stripes. 

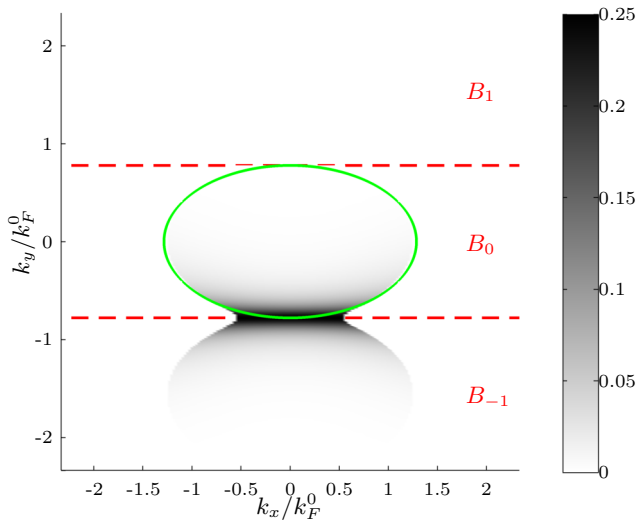
\begin{figure}[htb]
\def\svgwidth{.98\columnwidth}
\begingroup%
  \makeatletter%
  \providecommand\color[2][]{%
    \errmessage{(Inkscape) Color is used for the text in Inkscape, but the package 'color.sty' is not loaded}%
    \renewcommand\color[2][]{}%
  }%
  \providecommand\transparent[1]{%
    \errmessage{(Inkscape) Transparency is used (non-zero) for the text in Inkscape, but the package 'transparent.sty' is not loaded}%
    \renewcommand\transparent[1]{}%
  }%
  \providecommand\rotatebox[2]{#2}%
  \ifx\svgwidth\undefined%
    \setlength{\unitlength}{357.7504bp}%
    \ifx\svgscale\undefined%
      \relax%
    \else%
      \setlength{\unitlength}{\unitlength * \real{\svgscale}}%
    \fi%
  \else%
    \setlength{\unitlength}{\svgwidth}%
  \fi%
  \global\let\svgwidth\undefined%
  \global\let\svgscale\undefined%
  \makeatother%
  \begin{picture}(1,0.82389957)%
    \put(0,0){\includegraphics[width=\unitlength]{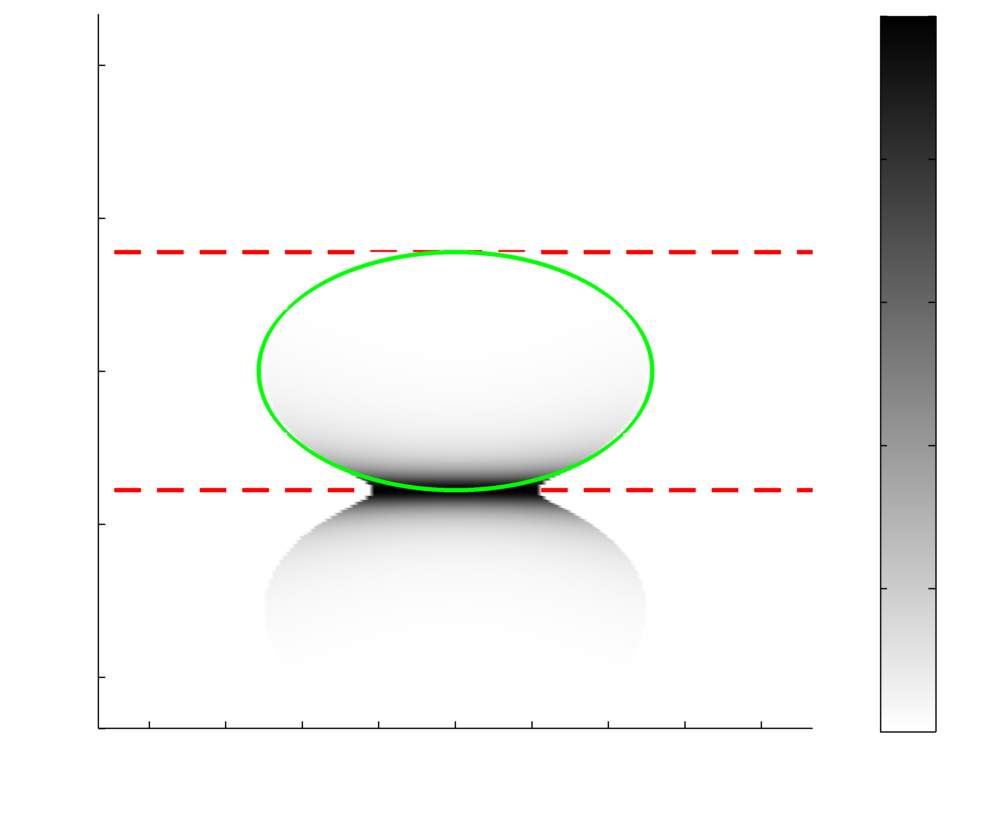}}%
    \put(0.45562268,0.02301943){\color[rgb]{0,0,0}\makebox(0,0)[b]{\smash{$k_x/\kfo$}}}%
    \put(0.04751469,0.44999867){\color[rgb]{0,0,0}\rotatebox{90}{\makebox(0,0)[b]{\smash{$k_y/\kfo$}}}}%
    \put(0.73280587,0.67515676){\color[rgb]{1,0,0}\makebox(0,0)[lb]{\smash{$B_1$}}}%
    \put(0.73280587,0.43595536){\color[rgb]{1,0,0}\makebox(0,0)[lb]{\smash{$B_0$}}}%
    \put(0.73280587,0.19675172){\color[rgb]{1,0,0}\makebox(0,0)[lb]{\smash{$B_{-1}$}}}%
    \put(0.15019186,0.0531544){\color[rgb]{0,0,0}\makebox(0,0)[b]{\smash{\scriptsize -2}}}%
    \put(0.22707451,0.0531544){\color[rgb]{0,0,0}\makebox(0,0)[b]{\smash{\scriptsize -1.5}}}%
    \put(0.30395716,0.0531544){\color[rgb]{0,0,0}\makebox(0,0)[b]{\smash{\scriptsize -1}}}%
    \put(0.38083981,0.0531544){\color[rgb]{0,0,0}\makebox(0,0)[b]{\smash{\scriptsize -0.5}}}%
    \put(0.45772023,0.0531544){\color[rgb]{0,0,0}\makebox(0,0)[b]{\smash{\scriptsize 0}}}%
    \put(0.53460288,0.0531544){\color[rgb]{0,0,0}\makebox(0,0)[b]{\smash{\scriptsize 0.5}}}%
    \put(0.61148553,0.0531544){\color[rgb]{0,0,0}\makebox(0,0)[b]{\smash{\scriptsize 1}}}%
    \put(0.68836818,0.0531544){\color[rgb]{0,0,0}\makebox(0,0)[b]{\smash{\scriptsize 1.5}}}%
    \put(0.76525083,0.0531544){\color[rgb]{0,0,0}\makebox(0,0)[b]{\smash{\scriptsize 2}}}%
    \put(0.08456398,0.1340868){\color[rgb]{0,0,0}\makebox(0,0)[rb]{\smash{\scriptsize -2}}}%
    \put(0.08456398,0.28784987){\color[rgb]{0,0,0}\makebox(0,0)[rb]{\smash{\scriptsize -1}}}%
    \put(0.08456398,0.44161517){\color[rgb]{0,0,0}\makebox(0,0)[rb]{\smash{\scriptsize 0}}}%
    \put(0.08456398,0.59537824){\color[rgb]{0,0,0}\makebox(0,0)[rb]{\smash{\scriptsize 1}}}%
    \put(0.08456398,0.74914354){\color[rgb]{0,0,0}\makebox(0,0)[rb]{\smash{\scriptsize 2}}}%
    \put(0.94108962,0.07924634){\color[rgb]{0,0,0}\makebox(0,0)[lb]{\smash{\scriptsize 0}}}%
    \put(0.94108962,0.22306057){\color[rgb]{0,0,0}\makebox(0,0)[lb]{\smash{\scriptsize 0.05}}}%
    \put(0.94108962,0.36687703){\color[rgb]{0,0,0}\makebox(0,0)[lb]{\smash{\scriptsize 0.1}}}%
    \put(0.94108962,0.5106935){\color[rgb]{0,0,0}\makebox(0,0)[lb]{\smash{\scriptsize 0.15}}}%
    \put(0.94108962,0.65450773){\color[rgb]{0,0,0}\makebox(0,0)[lb]{\smash{\scriptsize 0.2}}}%
    \put(0.94108962,0.79832196){\color[rgb]{0,0,0}\makebox(0,0)[lb]{\smash{\scriptsize 0.25}}}%
  \end{picture}%
\endgroup%
 \caption{(color online) The correlation function ${\cal C}(\kv,\kv+\qcv)$ for $g=1.01$, $\theta=0.3\pi$ in the first three Brillouin zones. It has a peak value of $1/4$ at $\kv=-\qcv/2$. Also shown is the Fermi surface for a homogenous phase with same tilting angle and interaction strength.}
 \label{fig:correlation}
\end{figure}

In TOF experiments, 
the density-density correlation function $\langle \rho(\rv) \rho(\rv')\rangle$ at points $\rv$ and $\rv'$ can be measured after the trap has been switched off and the gas has been 
allowed to expand for a time $t$. Assuming free expansion, this corresponds to measuring the momentum correlation function $\mean{n_{\kv}n_{\kpv}}$
before expansion with $\kv=m\rv/t$. We therefore analyse the correlation function
\begin{equation}
 \label{eq:correlation}
 {\cal C}(\kv,\kpv)=\mean{n_{\kv}n_{\kpv}}-\mean{n_{\kv}}\mean{n_{\kpv}}=- \mean{\cd{\kv}\co{\kpv}}\mean{\cd{\kpv}\co{\kv}},
\end{equation}
where we have used mean-field theory in the second equality and assumed $\kv\neq\kpv$. 
 ${\cal C}(\kv,\kpv)$ is nonzero in the striped phase for $\kpv=\kv\pm\qcv$, and taking $\kpv=\kv+\qcv$ we obtain 
\begin{widetext}
\begin{equation}
 \label{eq:correlation_qc}
 {\cal C}(\kv,\kv+\qcv)=\begin{cases}
0 &\textup{ for } \kv \in B_{1}\\
-\norm{\sum_{l=1}^3U({\kv})_{1,l}^{*}U({\kv})_{2,l}f(E_{\kv,l})}^2 &\textup{ for } \kv \in B_0 \\
-\norm{\sum_{l=1}^3U({\kv+\qcv})_{2,l}^{*}U({\kv}+\qcv)_{3,l}f(E_{\kv+\qcv,l})}^2 &\textup{ for } \kv \in B_{-1}.
\end{cases}
\end{equation}
\end{widetext}
In Fig.~\ref{fig:correlation}, we plot ${\cal C}(\kv,\kv+\qcv)$ as a function of $\kv$ for $g=1.01$, $\theta=0.3\pi$. 

We clearly see a peak around $\kv=-\qcv/2$, where the nearly degenerate states on opposite sites of the Fermi ellipse strongly mix. Right 
at the Fermi surface for $k_x=0$, the states are fully mixed giving the maximum value $\norm{{\cal C}(\kv,\kv+\qcv)}^2=1/4$. 
Note that ${\cal C}(\kv,\kv-\qcv)$ is the same as \eqref{eq:correlation_qc} mirrored around the $k_x$-axis.
We conclude that 
stripe order can be detected by the presence of characteristic peaks in the density-density correlation function measured in a TOF experiment. 

It should be noted that the dipolar interaction is long ranged compared to the usual atom-atom interaction, and it is therefore not obvious that it can be 
neglected during free expansion. However, we expect that the smoking gun features of the striped phase  are robust toward interaction effects during TOF. 
In particular, the peak in the correlation function shown in Fig.~\ref{fig:correlation} will survive interaction effects,
 since it is an inherent feature of the density wave. It might be distorted during expansion due to interactions, but it will not  disappear. Also, it was found  in Ref. \cite{Sogo2009} that for pancake shaped traps, which is the geometry we consider, dipolar interactions only have a small effect on the momentum distribution in TOF experiments. Even though they only analysed the case when the dipoles are perpendicular to the pancake, this indicates that interaction effects are small under TOF experiments for the system we consider.

\section{Discussion}
Since stripe formation  occurs for fairly strong coupling, one cannot  expect in general the mean-field approach used in this paper to be quantitatively accurate. 
From the 1D character of the stripes, we do indeed expect fluctuations away from the mean-field result to be significant. These fluctuations will suppress stripe order and
therefore increase the critical coupling strength for stripe formation. It is difficult to give a precise estimate of the accuracy of the mean-field approach. 
 From the Ginzburg or Brout criterion, we expect the size of fluctuations to be determined by $h(k)/\epsilon_F$.  
 
 One can also compare our prediction for the 
 critical coupling strength with those  obtained using different theoretical approaches. 
 Calculations of the divergence of the density-density response function using the RPA  give a critical coupling strength of 
 $g\simeq0.26$ for dipoles oriented perpendicular to the plane, $\theta=0$~\cite{Yamaguchi2010,Sun2010}. Including exchange correlations 
 to form a conserving Hartree-Fock approximation increases the critical coupling strength to about $g=0.57$~\cite{Sieberer2011,Babadi2011a,Block2012} which is equivalent to the approach in the 
present paper. When the response function is calculated  introducing a local field factor
 using the so called STLS scheme, one obtains $g=2.6$ for the critical coupling strength~\cite{Parish2012}.
This approach does include correlations beyond Hartree-Fock, but at the same time it neglects the non-local nature of the correlations. 
For the special case of the dipoles perpendicular to the plane,  
fixed-node  Monte-Carlo calculations indicate that the striped phase is preceded by a triangular Wigner crystal at 
$g=10.6\pm 1.3$~\cite{Matveeva2012}. However, these calculations use variational wave functions both for the normal and the striped phase with less 
than 100 particles, and the resulting energies of the two phases differ by less than 1\%. Thus, it is not clear how robust these results are to finite size effects and 
 to improvements in the wave functions. Finally, a variational method based on wave functions for the 2D electron gas yields  $g=11.9\pm 1.7$ for the critical 
coupling strength of the Wigner crystal~\cite{Babadi2013}, which is close to the diffusion Monte Carlo result, but to our knowledge 
this method has not been used to look for the density wave instability. In total, the  large discrepancies between the different theoretical 
predictions show that the striped phase is a strongly correlated phenomenon. It is unfortunately not straightforward to rank the accuracy of the different results  
approaches, and the problem therefore calls for an experimental investigation. 
We emphasize  that the analysis presented here based on the mean-field approach is the first one to describe the broken symmetry striped phase. Also, 
the mean-field results presented  here such as the smoking gun features of the striped phase in the momentum distribution, must be expected to  be qualitatively correct.

The striped phase spontaneously breaks translational symmetry, and it is therefore a quantum analogue of the  classical smectic phases~\cite{Chaikin1995}.
Quantum nematic and smectic phases play a significant role for many interesting electronic materials which have been discovered in the last couple of 
decades~\cite{Kivelson1998,Fradkin2010}.  
Contrary to the electron systems, which are plagued by  impurities, intricate band structures, 
lattice defects and distortions, that complicate a systematic analysis, ultracold dipolar gases are extremely clean and in addition experimentally very flexible. 
They therefore provide a great opportunity to investigate the formation of a smectic phase in a controlled and pure setting.
In this paper, we approach these phases from 
the weak-coupling perspective, where the phases arise from the successive breaking of symmetries of an underlying Fermi surface~\cite{Fradkin2010b}.

\section{Experimental considerations}
The interaction strength depends strongly on the type of molecule used in an experiment, since it scales with $g\propto m p^2 \sqrt{n_\textup{2D}}$. 
As an example,  consider the chemically stable ${}^{23}\textup{Na}^{40}\textup{K}$. To give a conservative estimate of experimentally realistic values, 
we refer to Refs.~\onlinecite{Ni2010,DeMiranda2011}, where the JILA group reports a maximum value of the induced dipole moment $p/p_0$ of about $0.4$ for the chemically unstable ${}^{40}\textup{K}^{87}\textup{Rb}$,
 and a maximum density of $3.4 \cdot 10^{7}\textup{ cm}^{-2}$ in a pancake geometry. If the same values are used for ${}^{23}\textup{Na}^{40}\textup{K}$, it corresponds
  to $g\simeq1.0$ which is well within the striped phase as calculated in  mean-field theory presented here. Similar parameters for the other chemically stable 
   molecule $^{40}\textup{K}^{133}\textup{Cs}$ gives $g=1.3$, and taking $p=p_0$ yields  $g=8.4$.
   Provided
    one can overcome difficulties related to the fact that $^{6}\textup{Li}^{133}\textup{Cs}$ is chemically reactive by using the 2D geometry, 
    the  large permanent dipole moment  $p_0=5.5$ Debye of this molecule means that one can even achieve the very high coupling strength  
    $g\simeq 55$ for the same density. These very different values illustrate the 
    quadratic dependence of the coupling strength on the dipole moment, which means that experiments likely will be able to 
    probe a large region of the phase diagram. One should therefore be able to investigate the critical value for 
    stripe formation, which is presently not settled theoretically as discussed above.

An interesting consideration is the effects of temperature. A finite and small temperature broadens the momentum distribution leading to less sharp signatures.
In the strict 2D limit, no true long range order exists at non-zero temperature, and the phase transition to the homogenous phase  is of the Berezinskii-Kosterlitz-Thouless (BKT) 
type~\cite{Berezinskii1972,Kosterlitz1973}. For increasing temperatures, defects in the form of insertion and disappearance of stripes will proliferate eventually melting the stripes. In the strongly interacting limit we expect the BKT temperature to be proportional to the density, however the constant of proportionality has yet to be calculated. This will be explored in future work. 

\section{Conclusions} 
We studied the $T=0$ properties of the striped phase of a 2D system of fermionic dipoles aligned by an external field. A Hartree-Fock 
theory was developed, which was shown to recover previous results for the critical coupling strength for stripe formation. The amplitude of the stripes was 
calculated as a function of the dipole moment and orientation, and the 
quasiparticle spectrum of the striped phase was shown to exhibit a 1D Brillouin zone structure with gapped 
as well as gapless regions around the Fermi surface. The system therefore remains compressible in the striped phase, and it collapses for essentially 
the same dipole strength as in the normal phase. We showed that the striped phase has clear signatures in the 
  momentum distribution and in the momentum correlations, which can both be measured in TOF experiments. Finally, we discussed how the striped phase 
 can be realised with experimentally relevant molecules. 

\section*{Acknowledgments}
We are grateful to Nikolaj Zinner for discussions and to  the Centre for Scientific Computing in Aarhus for computation time. GMB would like to acknowledge the support of the Carlsberg Foundation via grant 2011 01 0264 and the Villum Foundation via grant VKR023163.
\appendix
\section{$k$-space in the conserving Hartree-Fock approximation} \label{app:k-space}
To argue for the truncation of $k$-space, we examine the calculation of the static density-density response function $\chi$. The divergence of $\chi(\qv,\omega=0)$ signifies the instability of the system towards forming density waves with wave vector $\qv$~\cite{Yamaguchi2010,Sun2010} and thus marks the boundary of the DW phase. The self consistent mean-field theory employed in this study is an extension of the conserving\cite{Baym1961} HFA to the density-density response function as calculated in \cite{Sieberer2011,Babadi2011a,Block2012}. So examining the latter approach gives an indication of which $k$-states are relevant in the vicinity of the phase transition.
As shown in \cite{Block2012}, the internal Matsubara frequencies in the exchange plus direct interaction approximation to $\chi$ only appear in the particle-hole propagator $\Pi(k,q)=G(k+q)G(k)$, where $k=(\kv,ik_n)$ is the 2+1 momentum and $G$ is the fully dressed single particle Greens function. The Matsubara frequency sum is trivial so
\begin{align}
\label{eq:densityresponse}
& \sum_{k}\Pi(k,q)=\sum_{\kv} \frac{f_{\kv}-f_{\kv+\qv}}{iq_n+\epsilon_{\kv}-\epsilon_{\kv+\qv}}\\
 &=\sum_{\kv} \bigg( \frac{f_{\kv} }{iq_n+\epsilon_{\kv}-\epsilon_{\kv+\qv}} +\frac{f_{\kv} }{-iq_n+\epsilon_{\kv}-\epsilon_{\kv-\qv}} \bigg)
\end{align}
where $\epsilon_{\mathbf k}$ is the Hartree Fock single particle energy as given by \eqref{eq:epsilon}. Here we can see that particle-hole propagator is given exactly by the coupling between the occupied states of the lowest band $\epsilon_{\kv}$ to the two bands $\epsilon_{\kv\pm\qv}$ which is captured by the three band model described in section~\ref{sec:three-band-theory}.


\begin{thebibliography}{44}%
\makeatletter
\providecommand \@ifxundefined [1]{%
 \@ifx{#1\undefined}
}%
\providecommand \@ifnum [1]{%
 \ifnum #1\expandafter \@firstoftwo
 \else \expandafter \@secondoftwo
 \fi
}%
\providecommand \@ifx [1]{%
 \ifx #1\expandafter \@firstoftwo
 \else \expandafter \@secondoftwo
 \fi
}%
\providecommand \natexlab [1]{#1}%
\providecommand \enquote  [1]{``#1''}%
\providecommand \bibnamefont  [1]{#1}%
\providecommand \bibfnamefont [1]{#1}%
\providecommand \citenamefont [1]{#1}%
\providecommand \href@noop [0]{\@secondoftwo}%
\providecommand \href [0]{\begingroup \@sanitize@url \@href}%
\providecommand \@href[1]{\@@startlink{#1}\@@href}%
\providecommand \@@href[1]{\endgroup#1\@@endlink}%
\providecommand \@sanitize@url [0]{\catcode `\\12\catcode `\$12\catcode
  `\&12\catcode `\#12\catcode `\^12\catcode `\_12\catcode `\%12\relax}%
\providecommand \@@startlink[1]{}%
\providecommand \@@endlink[0]{}%
\providecommand \url  [0]{\begingroup\@sanitize@url \@url }%
\providecommand \@url [1]{\endgroup\@href {#1}{\urlprefix }}%
\providecommand \urlprefix  [0]{URL }%
\providecommand \Eprint [0]{\href }%
\providecommand \doibase [0]{http://dx.doi.org/}%
\providecommand \selectlanguage [0]{\@gobble}%
\providecommand \bibinfo  [0]{\@secondoftwo}%
\providecommand \bibfield  [0]{\@secondoftwo}%
\providecommand \translation [1]{[#1]}%
\providecommand \BibitemOpen [0]{}%
\providecommand \bibitemStop [0]{}%
\providecommand \bibitemNoStop [0]{.\EOS\space}%
\providecommand \EOS [0]{\spacefactor3000\relax}%
\providecommand \BibitemShut  [1]{\csname bibitem#1\endcsname}%
\let\auto@bib@innerbib\@empty
\bibitem [{\citenamefont {Bloch}\ \emph {et~al.}(2008)\citenamefont {Bloch},
  \citenamefont {Dalibard},\ and\ \citenamefont {Zwerger}}]{Bloch2008}%
  \BibitemOpen
  \bibfield  {author} {\bibinfo {author} {\bibfnamefont {I.}~\bibnamefont
  {Bloch}}, \bibinfo {author} {\bibfnamefont {J.}~\bibnamefont {Dalibard}}, \
  and\ \bibinfo {author} {\bibfnamefont {W.}~\bibnamefont {Zwerger}},\ }\href
  {\doibase 10.1103/RevModPhys.80.885} {\bibfield  {journal} {\bibinfo
  {journal} {Reviews of Modern Physics}\ }\textbf {\bibinfo {volume} {80}},\
  \bibinfo {pages} {885} (\bibinfo {year} {2008})}\BibitemShut {NoStop}%
\bibitem [{\citenamefont {Giorgini}\ \emph {et~al.}(2008)\citenamefont
  {Giorgini}, \citenamefont {Pitaevskii},\ and\ \citenamefont
  {Stringari}}]{Giorgini2008a}%
  \BibitemOpen
  \bibfield  {author} {\bibinfo {author} {\bibfnamefont {S.}~\bibnamefont
  {Giorgini}}, \bibinfo {author} {\bibfnamefont {L.~L.}\ \bibnamefont
  {Pitaevskii}}, \ and\ \bibinfo {author} {\bibfnamefont {S.}~\bibnamefont
  {Stringari}},\ }\href {\doibase doi: 10.1103/revmodphys.80.1215} {\bibfield
  {journal} {\bibinfo  {journal} {Reviews of Modern Physics}\ }\textbf
  {\bibinfo {volume} {80}},\ \bibinfo {pages} {1215} (\bibinfo {year}
  {2008})}\BibitemShut {NoStop}%
\bibitem [{\citenamefont {Ni}\ \emph {et~al.}(2010)\citenamefont {Ni},
  \citenamefont {Ospelkaus}, \citenamefont {Wang}, \citenamefont
  {Qu\'{e}m\'{e}ner}, \citenamefont {Neyenhuis}, \citenamefont {de~Miranda},
  \citenamefont {Bohn}, \citenamefont {Ye},\ and\ \citenamefont
  {Jin}}]{Ni2010}%
  \BibitemOpen
  \bibfield  {author} {\bibinfo {author} {\bibfnamefont {K.-K.}\ \bibnamefont
  {Ni}}, \bibinfo {author} {\bibfnamefont {S.}~\bibnamefont {Ospelkaus}},
  \bibinfo {author} {\bibfnamefont {D.}~\bibnamefont {Wang}}, \bibinfo {author}
  {\bibfnamefont {G.}~\bibnamefont {Qu\'{e}m\'{e}ner}}, \bibinfo {author}
  {\bibfnamefont {B.}~\bibnamefont {Neyenhuis}}, \bibinfo {author}
  {\bibfnamefont {M.~H.~G.}\ \bibnamefont {de~Miranda}}, \bibinfo {author}
  {\bibfnamefont {J.~L.}\ \bibnamefont {Bohn}}, \bibinfo {author}
  {\bibfnamefont {J.}~\bibnamefont {Ye}}, \ and\ \bibinfo {author}
  {\bibfnamefont {D.~S.}\ \bibnamefont {Jin}},\ }\href {\doibase
  10.1038/nature08953} {\bibfield  {journal} {\bibinfo  {journal} {Nature}\
  }\textbf {\bibinfo {volume} {464}},\ \bibinfo {pages} {1324} (\bibinfo {year}
  {2010})}\BibitemShut {NoStop}%
\bibitem [{\citenamefont {Deh}\ \emph {et~al.}(2010)\citenamefont {Deh},
  \citenamefont {Gunton}, \citenamefont {Klappauf}, \citenamefont {Li},
  \citenamefont {Semczuk}, \citenamefont {{Van Dongen}},\ and\ \citenamefont
  {Madison}}]{Deh2010}%
  \BibitemOpen
  \bibfield  {author} {\bibinfo {author} {\bibfnamefont {B.}~\bibnamefont
  {Deh}}, \bibinfo {author} {\bibfnamefont {W.}~\bibnamefont {Gunton}},
  \bibinfo {author} {\bibfnamefont {B.~G.}\ \bibnamefont {Klappauf}}, \bibinfo
  {author} {\bibfnamefont {Z.}~\bibnamefont {Li}}, \bibinfo {author}
  {\bibfnamefont {M.}~\bibnamefont {Semczuk}}, \bibinfo {author} {\bibfnamefont
  {J.}~\bibnamefont {{Van Dongen}}}, \ and\ \bibinfo {author} {\bibfnamefont
  {K.~W.}\ \bibnamefont {Madison}},\ }\href {\doibase
  10.1103/PhysRevA.82.020701} {\bibfield  {journal} {\bibinfo  {journal}
  {Physical Review A}\ }\textbf {\bibinfo {volume} {82}},\ \bibinfo {pages}
  {020701} (\bibinfo {year} {2010})}\BibitemShut {NoStop}%
\bibitem [{\citenamefont {Heo}\ \emph {et~al.}(2012)\citenamefont {Heo},
  \citenamefont {Wang}, \citenamefont {Christensen}, \citenamefont {Rvachov},
  \citenamefont {Cotta}, \citenamefont {Choi}, \citenamefont {Lee},\ and\
  \citenamefont {Ketterle}}]{Heo2012}%
  \BibitemOpen
  \bibfield  {author} {\bibinfo {author} {\bibfnamefont {M.-S.}\ \bibnamefont
  {Heo}}, \bibinfo {author} {\bibfnamefont {T.~T.}\ \bibnamefont {Wang}},
  \bibinfo {author} {\bibfnamefont {C.~A.}\ \bibnamefont {Christensen}},
  \bibinfo {author} {\bibfnamefont {T.~M.}\ \bibnamefont {Rvachov}}, \bibinfo
  {author} {\bibfnamefont {D.~A.}\ \bibnamefont {Cotta}}, \bibinfo {author}
  {\bibfnamefont {J.-H.}\ \bibnamefont {Choi}}, \bibinfo {author}
  {\bibfnamefont {Y.-R.}\ \bibnamefont {Lee}}, \ and\ \bibinfo {author}
  {\bibfnamefont {W.}~\bibnamefont {Ketterle}},\ }\href {\doibase
  10.1103/PhysRevA.86.021602} {\bibfield  {journal} {\bibinfo  {journal}
  {Physical Review A}\ }\textbf {\bibinfo {volume} {86}},\ \bibinfo {pages}
  {021602} (\bibinfo {year} {2012})}\BibitemShut {NoStop}%
\bibitem [{\citenamefont {Wu}\ \emph {et~al.}(2012)\citenamefont {Wu},
  \citenamefont {Park}, \citenamefont {Ahmadi}, \citenamefont {Will},\ and\
  \citenamefont {Zwierlein}}]{Wu2012}%
  \BibitemOpen
  \bibfield  {author} {\bibinfo {author} {\bibfnamefont {C.-H.}\ \bibnamefont
  {Wu}}, \bibinfo {author} {\bibfnamefont {J.~W.}\ \bibnamefont {Park}},
  \bibinfo {author} {\bibfnamefont {P.}~\bibnamefont {Ahmadi}}, \bibinfo
  {author} {\bibfnamefont {S.}~\bibnamefont {Will}}, \ and\ \bibinfo {author}
  {\bibfnamefont {M.~W.}\ \bibnamefont {Zwierlein}},\ }\href {\doibase
  10.1103/PhysRevLett.109.085301} {\bibfield  {journal} {\bibinfo  {journal}
  {Phys. Rev. Lett.}\ }\textbf {\bibinfo {volume} {109}},\ \bibinfo {pages}
  {085301} (\bibinfo {year} {2012})}\BibitemShut {NoStop}%
\bibitem [{\citenamefont {Schulze}\ \emph {et~al.}(2013)\citenamefont
  {Schulze}, \citenamefont {Temelkov}, \citenamefont {Gempel}, \citenamefont
  {Hartmann}, \citenamefont {Kn\"{o}ckel}, \citenamefont {Ospelkaus},\ and\
  \citenamefont {Tiemann}}]{Schulze2013}%
  \BibitemOpen
  \bibfield  {author} {\bibinfo {author} {\bibfnamefont {T.~A.}\ \bibnamefont
  {Schulze}}, \bibinfo {author} {\bibfnamefont {I.~I.}\ \bibnamefont
  {Temelkov}}, \bibinfo {author} {\bibfnamefont {M.~W.}\ \bibnamefont
  {Gempel}}, \bibinfo {author} {\bibfnamefont {T.}~\bibnamefont {Hartmann}},
  \bibinfo {author} {\bibfnamefont {H.}~\bibnamefont {Kn\"{o}ckel}}, \bibinfo
  {author} {\bibfnamefont {S.}~\bibnamefont {Ospelkaus}}, \ and\ \bibinfo
  {author} {\bibfnamefont {E.}~\bibnamefont {Tiemann}},\ }\href {\doibase
  10.1103/PhysRevA.88.023401} {\bibfield  {journal} {\bibinfo  {journal}
  {Physical Review A}\ }\textbf {\bibinfo {volume} {88}},\ \bibinfo {pages}
  {023401} (\bibinfo {year} {2013})}\BibitemShut {NoStop}%
\bibitem [{\citenamefont {Tung}\ \emph {et~al.}(2013)\citenamefont {Tung},
  \citenamefont {Parker}, \citenamefont {Johansen}, \citenamefont {Chin},
  \citenamefont {Wang},\ and\ \citenamefont {Julienne}}]{Tung2013}%
  \BibitemOpen
  \bibfield  {author} {\bibinfo {author} {\bibfnamefont {S.-K.}\ \bibnamefont
  {Tung}}, \bibinfo {author} {\bibfnamefont {C.}~\bibnamefont {Parker}},
  \bibinfo {author} {\bibfnamefont {J.}~\bibnamefont {Johansen}}, \bibinfo
  {author} {\bibfnamefont {C.}~\bibnamefont {Chin}}, \bibinfo {author}
  {\bibfnamefont {Y.}~\bibnamefont {Wang}}, \ and\ \bibinfo {author}
  {\bibfnamefont {P.~S.}\ \bibnamefont {Julienne}},\ }\href {\doibase
  10.1103/PhysRevA.87.010702} {\bibfield  {journal} {\bibinfo  {journal}
  {Physical Review A}\ }\textbf {\bibinfo {volume} {87}},\ \bibinfo {pages}
  {010702} (\bibinfo {year} {2013})}\BibitemShut {NoStop}%
\bibitem [{\citenamefont {Repp}\ \emph {et~al.}(2013)\citenamefont {Repp},
  \citenamefont {Pires}, \citenamefont {Ulmanis}, \citenamefont {Heck},
  \citenamefont {Kuhnle}, \citenamefont {Weidem\"{u}ller},\ and\ \citenamefont
  {Tiemann}}]{Repp2013}%
  \BibitemOpen
  \bibfield  {author} {\bibinfo {author} {\bibfnamefont {M.}~\bibnamefont
  {Repp}}, \bibinfo {author} {\bibfnamefont {R.}~\bibnamefont {Pires}},
  \bibinfo {author} {\bibfnamefont {J.}~\bibnamefont {Ulmanis}}, \bibinfo
  {author} {\bibfnamefont {R.}~\bibnamefont {Heck}}, \bibinfo {author}
  {\bibfnamefont {E.~D.}\ \bibnamefont {Kuhnle}}, \bibinfo {author}
  {\bibfnamefont {M.}~\bibnamefont {Weidem\"{u}ller}}, \ and\ \bibinfo {author}
  {\bibfnamefont {E.}~\bibnamefont {Tiemann}},\ }\href {\doibase
  10.1103/PhysRevA.87.010701} {\bibfield  {journal} {\bibinfo  {journal}
  {Physical Review A}\ }\textbf {\bibinfo {volume} {87}},\ \bibinfo {pages}
  {010701} (\bibinfo {year} {2013})}\BibitemShut {NoStop}%
\bibitem [{\citenamefont {Lahaye}\ \emph {et~al.}(2009)\citenamefont {Lahaye},
  \citenamefont {Menotti}, \citenamefont {Santos}, \citenamefont {Lewenstein},\
  and\ \citenamefont {Pfau}}]{Lahaye2009}%
  \BibitemOpen
  \bibfield  {author} {\bibinfo {author} {\bibfnamefont {T.}~\bibnamefont
  {Lahaye}}, \bibinfo {author} {\bibfnamefont {C.}~\bibnamefont {Menotti}},
  \bibinfo {author} {\bibfnamefont {L.}~\bibnamefont {Santos}}, \bibinfo
  {author} {\bibfnamefont {M.}~\bibnamefont {Lewenstein}}, \ and\ \bibinfo
  {author} {\bibfnamefont {T.}~\bibnamefont {Pfau}},\ }\href {\doibase doi:
  10.1088/0034-4885/72/12/126401} {\bibfield  {journal} {\bibinfo  {journal}
  {Reports on Progress in Physics}\ }\textbf {\bibinfo {volume} {72}},\
  \bibinfo {pages} {126401} (\bibinfo {year} {2009})}\BibitemShut {NoStop}%
\bibitem [{\citenamefont {de~Miranda}\ \emph {et~al.}(2011)\citenamefont
  {de~Miranda}, \citenamefont {Chotia}, \citenamefont {Neyenhuis},
  \citenamefont {Wang}, \citenamefont {Quemener}, \citenamefont {Ospelkaus},
  \citenamefont {Bohn}, \citenamefont {Ye},\ and\ \citenamefont
  {Jin}}]{DeMiranda2011}%
  \BibitemOpen
  \bibfield  {author} {\bibinfo {author} {\bibfnamefont {M.~H.~G.}\
  \bibnamefont {de~Miranda}}, \bibinfo {author} {\bibfnamefont
  {A.}~\bibnamefont {Chotia}}, \bibinfo {author} {\bibfnamefont
  {B.}~\bibnamefont {Neyenhuis}}, \bibinfo {author} {\bibfnamefont
  {D.}~\bibnamefont {Wang}}, \bibinfo {author} {\bibfnamefont {G.}~\bibnamefont
  {Quemener}}, \bibinfo {author} {\bibfnamefont {S.}~\bibnamefont {Ospelkaus}},
  \bibinfo {author} {\bibfnamefont {J.~L.}\ \bibnamefont {Bohn}}, \bibinfo
  {author} {\bibfnamefont {J.}~\bibnamefont {Ye}}, \ and\ \bibinfo {author}
  {\bibfnamefont {D.~S.}\ \bibnamefont {Jin}},\ }\href {\doibase doi:
  10.1038/nphys1939} {\bibfield  {journal} {\bibinfo  {journal} {Nature
  Physics}\ }\textbf {\bibinfo {volume} {7}},\ \bibinfo {pages} {502} (\bibinfo
  {year} {2011})}\BibitemShut {NoStop}%
\bibitem [{\citenamefont {Chotia}\ \emph {et~al.}(2012)\citenamefont {Chotia},
  \citenamefont {Neyenhuis}, \citenamefont {Moses}, \citenamefont {Yan},
  \citenamefont {Covey}, \citenamefont {Foss-Feig}, \citenamefont {Rey},
  \citenamefont {Jin},\ and\ \citenamefont {Ye}}]{Chotia2012}%
  \BibitemOpen
  \bibfield  {author} {\bibinfo {author} {\bibfnamefont {A.}~\bibnamefont
  {Chotia}}, \bibinfo {author} {\bibfnamefont {B.}~\bibnamefont {Neyenhuis}},
  \bibinfo {author} {\bibfnamefont {S.~A.}\ \bibnamefont {Moses}}, \bibinfo
  {author} {\bibfnamefont {B.}~\bibnamefont {Yan}}, \bibinfo {author}
  {\bibfnamefont {J.~P.}\ \bibnamefont {Covey}}, \bibinfo {author}
  {\bibfnamefont {M.}~\bibnamefont {Foss-Feig}}, \bibinfo {author}
  {\bibfnamefont {A.~M.}\ \bibnamefont {Rey}}, \bibinfo {author} {\bibfnamefont
  {D.~S.}\ \bibnamefont {Jin}}, \ and\ \bibinfo {author} {\bibfnamefont
  {J.}~\bibnamefont {Ye}},\ }\href {\doibase 10.1103/PhysRevLett.108.080405}
  {\bibfield  {journal} {\bibinfo  {journal} {Physical Review Letters}\
  }\textbf {\bibinfo {volume} {108}},\ \bibinfo {pages} {080405} (\bibinfo
  {year} {2012})}\BibitemShut {NoStop}%
\bibitem [{\citenamefont {Bruun}\ and\ \citenamefont
  {Taylor}(2008)}]{Bruun2008}%
  \BibitemOpen
  \bibfield  {author} {\bibinfo {author} {\bibfnamefont {G.~M.}\ \bibnamefont
  {Bruun}}\ and\ \bibinfo {author} {\bibfnamefont {E.}~\bibnamefont {Taylor}},\
  }\href {\doibase doi: 10.1103/physrevlett.101.245301} {\bibfield  {journal}
  {\bibinfo  {journal} {Physical Review Letters}\ }\textbf {\bibinfo {volume}
  {101}},\ \bibinfo {pages} {245301} (\bibinfo {year} {2008})}\BibitemShut
  {NoStop}%
\bibitem [{\citenamefont {Cooper}\ and\ \citenamefont
  {Shlyapnikov}(2009)}]{Cooper2009}%
  \BibitemOpen
  \bibfield  {author} {\bibinfo {author} {\bibfnamefont {N.~R.}\ \bibnamefont
  {Cooper}}\ and\ \bibinfo {author} {\bibfnamefont {G.~V.}\ \bibnamefont
  {Shlyapnikov}},\ }\href {\doibase 10.1103/PhysRevLett.103.155302} {\bibfield
  {journal} {\bibinfo  {journal} {Physical Review Letters}\ }\textbf {\bibinfo
  {volume} {103}},\ \bibinfo {pages} {155302} (\bibinfo {year}
  {2009})}\BibitemShut {NoStop}%
\bibitem [{\citenamefont {Fregoso}\ \emph {et~al.}(2009)\citenamefont
  {Fregoso}, \citenamefont {Sun}, \citenamefont {Fradkin},\ and\ \citenamefont
  {Lev}}]{Fregoso2009}%
  \BibitemOpen
  \bibfield  {author} {\bibinfo {author} {\bibfnamefont {B.~M.}\ \bibnamefont
  {Fregoso}}, \bibinfo {author} {\bibfnamefont {K.}~\bibnamefont {Sun}},
  \bibinfo {author} {\bibfnamefont {E.}~\bibnamefont {Fradkin}}, \ and\
  \bibinfo {author} {\bibfnamefont {B.~L.}\ \bibnamefont {Lev}},\ }\href
  {http://stacks.iop.org/1367-2630/11/i=10/a=103003} {\bibfield  {journal}
  {\bibinfo  {journal} {New Journal of Physics}\ }\textbf {\bibinfo {volume}
  {11}},\ \bibinfo {pages} {103003} (\bibinfo {year} {2009})}\BibitemShut
  {NoStop}%
\bibitem [{\citenamefont {Sun}\ \emph {et~al.}(2010)\citenamefont {Sun},
  \citenamefont {Wu},\ and\ \citenamefont {{Das Sarma}}}]{Sun2010}%
  \BibitemOpen
  \bibfield  {author} {\bibinfo {author} {\bibfnamefont {K.}~\bibnamefont
  {Sun}}, \bibinfo {author} {\bibfnamefont {C.}~\bibnamefont {Wu}}, \ and\
  \bibinfo {author} {\bibfnamefont {S.}~\bibnamefont {{Das Sarma}}},\ }\href
  {\doibase 10.1103/PhysRevB.82.075105} {\bibfield  {journal} {\bibinfo
  {journal} {Physical Review B}\ }\textbf {\bibinfo {volume} {82}},\ \bibinfo
  {pages} {075105} (\bibinfo {year} {2010})}\BibitemShut {NoStop}%
\bibitem [{\citenamefont {Yamaguchi}\ \emph {et~al.}(2010)\citenamefont
  {Yamaguchi}, \citenamefont {Sogo}, \citenamefont {Ito},\ and\ \citenamefont
  {Miyakawa}}]{Yamaguchi2010}%
  \BibitemOpen
  \bibfield  {author} {\bibinfo {author} {\bibfnamefont {Y.}~\bibnamefont
  {Yamaguchi}}, \bibinfo {author} {\bibfnamefont {T.}~\bibnamefont {Sogo}},
  \bibinfo {author} {\bibfnamefont {T.}~\bibnamefont {Ito}}, \ and\ \bibinfo
  {author} {\bibfnamefont {T.}~\bibnamefont {Miyakawa}},\ }\href {\doibase doi:
  10.1103/physreva.82.013643} {\bibfield  {journal} {\bibinfo  {journal}
  {Physical Review A}\ }\textbf {\bibinfo {volume} {82}},\ \bibinfo {pages}
  {013643} (\bibinfo {year} {2010})}\BibitemShut {NoStop}%
\bibitem [{\citenamefont {Zinner}\ and\ \citenamefont
  {Bruun}(2011)}]{Zinner2011}%
  \BibitemOpen
  \bibfield  {author} {\bibinfo {author} {\bibfnamefont {N.~T.}\ \bibnamefont
  {Zinner}}\ and\ \bibinfo {author} {\bibfnamefont {G.~M.}\ \bibnamefont
  {Bruun}},\ }\href {\doibase 10.1140/epjd/e2011-20094-3} {\bibfield  {journal}
  {\bibinfo  {journal} {The European Physical Journal D}\ }\textbf {\bibinfo
  {volume} {65}},\ \bibinfo {pages} {133} (\bibinfo {year} {2011})}\BibitemShut
  {NoStop}%
\bibitem [{\citenamefont {Babadi}\ and\ \citenamefont
  {Demler}(2011{\natexlab{a}})}]{Babadi2011}%
  \BibitemOpen
  \bibfield  {author} {\bibinfo {author} {\bibfnamefont {M.}~\bibnamefont
  {Babadi}}\ and\ \bibinfo {author} {\bibfnamefont {E.}~\bibnamefont
  {Demler}},\ }\href {\doibase doi: 10.1103/physreva.84.033636} {\bibfield
  {journal} {\bibinfo  {journal} {Physical Review A}\ }\textbf {\bibinfo
  {volume} {84}},\ \bibinfo {pages} {033636} (\bibinfo {year}
  {2011}{\natexlab{a}})}\BibitemShut {NoStop}%
\bibitem [{\citenamefont {Sieberer}\ and\ \citenamefont
  {Baranov}(2011)}]{Sieberer2011}%
  \BibitemOpen
  \bibfield  {author} {\bibinfo {author} {\bibfnamefont {L.~M.}\ \bibnamefont
  {Sieberer}}\ and\ \bibinfo {author} {\bibfnamefont {M.~A.}\ \bibnamefont
  {Baranov}},\ }\href {\doibase doi: 10.1103/physreva.84.063633} {\bibfield
  {journal} {\bibinfo  {journal} {Physical Review A}\ }\textbf {\bibinfo
  {volume} {84}},\ \bibinfo {pages} {063633} (\bibinfo {year}
  {2011})}\BibitemShut {NoStop}%
\bibitem [{\citenamefont {Parish}\ and\ \citenamefont
  {Marchetti}(2012)}]{Parish2012}%
  \BibitemOpen
  \bibfield  {author} {\bibinfo {author} {\bibfnamefont {M.~M.}\ \bibnamefont
  {Parish}}\ and\ \bibinfo {author} {\bibfnamefont {F.~M.}\ \bibnamefont
  {Marchetti}},\ }\href {\doibase 10.1103/PhysRevLett.108.145304} {\bibfield
  {journal} {\bibinfo  {journal} {Physical Review Letters}\ }\textbf {\bibinfo
  {volume} {108}},\ \bibinfo {pages} {145304} (\bibinfo {year}
  {2012})}\BibitemShut {NoStop}%
\bibitem [{\citenamefont {Block}\ \emph {et~al.}(2012)\citenamefont {Block},
  \citenamefont {Zinner},\ and\ \citenamefont {Bruun}}]{Block2012}%
  \BibitemOpen
  \bibfield  {author} {\bibinfo {author} {\bibfnamefont {J.~K.}\ \bibnamefont
  {Block}}, \bibinfo {author} {\bibfnamefont {N.~T.}\ \bibnamefont {Zinner}}, \
  and\ \bibinfo {author} {\bibfnamefont {G.~M.}\ \bibnamefont {Bruun}},\ }\href
  {\doibase 10.1088/1367-2630/14/10/105006} {\bibfield  {journal} {\bibinfo
  {journal} {New Journal of Physics}\ }\textbf {\bibinfo {volume} {14}},\
  \bibinfo {pages} {105006} (\bibinfo {year} {2012})}\BibitemShut {NoStop}%
\bibitem [{\citenamefont {Marchetti}\ and\ \citenamefont
  {Parish}(2013)}]{Marchetti2013}%
  \BibitemOpen
  \bibfield  {author} {\bibinfo {author} {\bibfnamefont {F.~M.}\ \bibnamefont
  {Marchetti}}\ and\ \bibinfo {author} {\bibfnamefont {M.~M.}\ \bibnamefont
  {Parish}},\ }\href {\doibase doi: 10.1103/physrevb.87.045110} {\bibfield
  {journal} {\bibinfo  {journal} {Physical Review B}\ }\textbf {\bibinfo
  {volume} {87}},\ \bibinfo {pages} {045110} (\bibinfo {year}
  {2013})}\BibitemShut {NoStop}%
\bibitem [{\citenamefont {Bruun}\ and\ \citenamefont
  {Nelson}(2014)}]{Bruun2014}%
  \BibitemOpen
  \bibfield  {author} {\bibinfo {author} {\bibfnamefont {G.~M.}\ \bibnamefont
  {Bruun}}\ and\ \bibinfo {author} {\bibfnamefont {D.~R.}\ \bibnamefont
  {Nelson}},\ }\href {\doibase 10.1103/PhysRevB.89.094112} {\bibfield
  {journal} {\bibinfo  {journal} {Physical Review B}\ }\textbf {\bibinfo
  {volume} {89}},\ \bibinfo {pages} {094112} (\bibinfo {year}
  {2014})}\BibitemShut {NoStop}%
\bibitem [{\citenamefont {{Lechner}}\ \emph {et~al.}(2014)\citenamefont
  {{Lechner}}, \citenamefont {{B{\"u}chler}},\ and\ \citenamefont
  {{Zoller}}}]{Lechner2014}%
  \BibitemOpen
  \bibfield  {author} {\bibinfo {author} {\bibfnamefont {W.}~\bibnamefont
  {{Lechner}}}, \bibinfo {author} {\bibfnamefont {H.-P.}\ \bibnamefont
  {{B{\"u}chler}}}, \ and\ \bibinfo {author} {\bibfnamefont {P.}~\bibnamefont
  {{Zoller}}},\ }\href@noop {} {\bibfield  {journal} {\bibinfo  {journal}
  {ArXiv e-prints}\ } (\bibinfo {year} {2014})},\ \Eprint
  {http://arxiv.org/abs/1401.5682} {arXiv:1401.5682 [cond-mat.stat-mech]}
  \BibitemShut {NoStop}%
\bibitem [{\citenamefont {Kivelson}\ \emph {et~al.}(1998)\citenamefont
  {Kivelson}, \citenamefont {Fradkin},\ and\ \citenamefont
  {Emery}}]{Kivelson1998}%
  \BibitemOpen
  \bibfield  {author} {\bibinfo {author} {\bibfnamefont {S.~A.}\ \bibnamefont
  {Kivelson}}, \bibinfo {author} {\bibfnamefont {E.}~\bibnamefont {Fradkin}}, \
  and\ \bibinfo {author} {\bibfnamefont {V.~J.}\ \bibnamefont {Emery}},\ }\href
  {\doibase 10.1038/31177} {\bibfield  {journal} {\bibinfo  {journal} {Nature}\
  }\textbf {\bibinfo {volume} {393}},\ \bibinfo {pages} {550} (\bibinfo {year}
  {1998})}\BibitemShut {NoStop}%
\bibitem [{\citenamefont {Fradkin}\ and\ \citenamefont
  {Kivelson}(2010)}]{Fradkin2010}%
  \BibitemOpen
  \bibfield  {author} {\bibinfo {author} {\bibfnamefont {E.}~\bibnamefont
  {Fradkin}}\ and\ \bibinfo {author} {\bibfnamefont {S.~A.}\ \bibnamefont
  {Kivelson}},\ }\href {\doibase 10.1126/science.1183464} {\bibfield  {journal}
  {\bibinfo  {journal} {Science}\ }\textbf {\bibinfo {volume} {327}},\ \bibinfo
  {pages} {155} (\bibinfo {year} {2010})}\BibitemShut {NoStop}%
\bibitem [{\citenamefont {Matveeva}\ and\ \citenamefont
  {Giorgini}(2012)}]{Matveeva2012}%
  \BibitemOpen
  \bibfield  {author} {\bibinfo {author} {\bibfnamefont {N.}~\bibnamefont
  {Matveeva}}\ and\ \bibinfo {author} {\bibfnamefont {S.}~\bibnamefont
  {Giorgini}},\ }\href {\doibase doi: 10.1103/physrevlett.109.200401}
  {\bibfield  {journal} {\bibinfo  {journal} {Physical Review Letters}\
  }\textbf {\bibinfo {volume} {109}},\ \bibinfo {pages} {200401} (\bibinfo
  {year} {2012})}\BibitemShut {NoStop}%
\bibitem [{\citenamefont {Baym}\ and\ \citenamefont
  {Kadanoff}(1961)}]{Baym1961}%
  \BibitemOpen
  \bibfield  {author} {\bibinfo {author} {\bibfnamefont {G.}~\bibnamefont
  {Baym}}\ and\ \bibinfo {author} {\bibfnamefont {L.}~\bibnamefont
  {Kadanoff}},\ }\href {\doibase 10.1103/PhysRev.124.287} {\bibfield  {journal}
  {\bibinfo  {journal} {Physical Review}\ }\textbf {\bibinfo {volume} {124}},\
  \bibinfo {pages} {287} (\bibinfo {year} {1961})}\BibitemShut {NoStop}%
\bibitem [{\citenamefont {Fischer}(2006)}]{Fischer2006}%
  \BibitemOpen
  \bibfield  {author} {\bibinfo {author} {\bibfnamefont {U.~R.}~\bibnamefont
  {Fischer}},\ }\href {\doibase 10.1103/PhysRevA.73.031602} {\bibfield
  {journal} {\bibinfo  {journal} {Physical Review A}\ }\textbf {\bibinfo
  {volume} {73}},\ \bibinfo {pages} {031602} (\bibinfo {year}
  {2006})}\BibitemShut {NoStop}%
\bibitem [{\citenamefont {Miyakawa}\ \emph {et~al.}(2008)\citenamefont
  {Miyakawa}, \citenamefont {Sogo},\ and\ \citenamefont {Pu}}]{Miyakawa2008}%
  \BibitemOpen
  \bibfield  {author} {\bibinfo {author} {\bibfnamefont {T.}~\bibnamefont
  {Miyakawa}}, \bibinfo {author} {\bibfnamefont {T.}~\bibnamefont {Sogo}}, \
  and\ \bibinfo {author} {\bibfnamefont {H.}~\bibnamefont {Pu}},\ }\href
  {\doibase doi: 10.1103/physreva.77.061603} {\bibfield  {journal} {\bibinfo
  {journal} {Physical Review A}\ }\textbf {\bibinfo {volume} {77}},\ \bibinfo
  {pages} {061603} (\bibinfo {year} {2008})}\BibitemShut {NoStop}%
\bibitem [{\citenamefont {Babadi}\ and\ \citenamefont
  {Demler}(2011{\natexlab{b}})}]{Babadi2011a}%
  \BibitemOpen
  \bibfield  {author} {\bibinfo {author} {\bibfnamefont {M.}~\bibnamefont
  {Babadi}}\ and\ \bibinfo {author} {\bibfnamefont {E.}~\bibnamefont
  {Demler}},\ }\href {\doibase doi: 10.1103/physrevb.84.235124} {\bibfield
  {journal} {\bibinfo  {journal} {Physical Review B}\ }\textbf {\bibinfo
  {volume} {84}},\ \bibinfo {pages} {235124} (\bibinfo {year}
  {2011}{\natexlab{b}})}\BibitemShut {NoStop}%
\bibitem [{\citenamefont {Chaikin}\ and\ \citenamefont
  {Lubensky}(1995)}]{Chaikin1995}%
  \BibitemOpen
  \bibfield  {author} {\bibinfo {author} {\bibfnamefont {P.~M.}\ \bibnamefont
  {Chaikin}}\ and\ \bibinfo {author} {\bibfnamefont {T.~C.}\ \bibnamefont
  {Lubensky}},\ }\href@noop {} {\emph {\bibinfo {title} {Principles of
  Condensed Matter Physics}}}\ (\bibinfo  {publisher} {Cambridge University
  Press},\ \bibinfo {year} {1995})\BibitemShut {NoStop}%
\bibitem [{\citenamefont {Andreev}\ and\ \citenamefont
  {Lifshitz}(1969)}]{Andreev1969}%
  \BibitemOpen
  \bibfield  {author} {\bibinfo {author} {\bibfnamefont {A.}~\bibnamefont
  {Andreev}}\ and\ \bibinfo {author} {\bibfnamefont {I.}~\bibnamefont
  {Lifshitz}},\ }\href@noop {} {\bibfield  {journal} {\bibinfo  {journal}
  {Sov.Phys.-JETP}\ }\textbf {\bibinfo {volume} {29}},\ \bibinfo {pages} {1107}
  (\bibinfo {year} {1969})}\BibitemShut {NoStop}%
\bibitem [{\citenamefont {Chester}(1970)}]{Chester1970}%
  \BibitemOpen
  \bibfield  {author} {\bibinfo {author} {\bibfnamefont {G.~V.}\ \bibnamefont
  {Chester}},\ }\href {\doibase 10.1103/PhysRevA.2.256} {\bibfield  {journal}
  {\bibinfo  {journal} {Physical Review A}\ }\textbf {\bibinfo {volume} {2}},\
  \bibinfo {pages} {256} (\bibinfo {year} {1970})}\BibitemShut {NoStop}%
\bibitem [{\citenamefont {Leggett}(1970)}]{Leggett1970}%
  \BibitemOpen
  \bibfield  {author} {\bibinfo {author} {\bibfnamefont {A.~J.}\ \bibnamefont
  {Leggett}},\ }\href {\doibase 10.1103/PhysRevLett.25.1543} {\bibfield
  {journal} {\bibinfo  {journal} {Physical Review Letters}\ }\textbf {\bibinfo
  {volume} {25}},\ \bibinfo {pages} {1543} (\bibinfo {year}
  {1970})}\BibitemShut {NoStop}%
\bibitem [{\citenamefont {Kim}\ and\ \citenamefont {Zubarev}(2004)}]{Kim2004}%
  \BibitemOpen
  \bibfield  {author} {\bibinfo {author} {\bibfnamefont {Y.~E.}\ \bibnamefont
  {Kim}}\ and\ \bibinfo {author} {\bibfnamefont {A.~L.}\ \bibnamefont
  {Zubarev}},\ }\href {\doibase 10.1103/PhysRevA.70.033612} {\bibfield
  {journal} {\bibinfo  {journal} {Materials Science}\ }\textbf {\bibinfo
  {volume} {70}},\ \bibinfo {pages} {1} (\bibinfo {year} {2004})}\BibitemShut
  {NoStop}%
\bibitem [{\citenamefont {Chan}(2008)}]{Chan2008}%
  \BibitemOpen
  \bibfield  {author} {\bibinfo {author} {\bibfnamefont {M.~H.~W.}\
  \bibnamefont {Chan}},\ }\href {\doibase 10.1126/science.1155302} {\bibfield
  {journal} {\bibinfo  {journal} {Science}\ }\textbf {\bibinfo {volume}
  {319}},\ \bibinfo {pages} {1207} (\bibinfo {year} {2008})}\BibitemShut
  {NoStop}%
\bibitem [{\citenamefont {Capogrosso-Sansone}\ \emph
  {et~al.}(2010)\citenamefont {Capogrosso-Sansone}, \citenamefont {Trefzger},
  \citenamefont {Lewenstein}, \citenamefont {Zoller},\ and\ \citenamefont
  {Pupillo}}]{Capogrosso-Sansone2010}%
  \BibitemOpen
  \bibfield  {author} {\bibinfo {author} {\bibfnamefont {B.}~\bibnamefont
  {Capogrosso-Sansone}}, \bibinfo {author} {\bibfnamefont {C.}~\bibnamefont
  {Trefzger}}, \bibinfo {author} {\bibfnamefont {M.}~\bibnamefont
  {Lewenstein}}, \bibinfo {author} {\bibfnamefont {P.}~\bibnamefont {Zoller}},
  \ and\ \bibinfo {author} {\bibfnamefont {G.}~\bibnamefont {Pupillo}},\ }\href
  {\doibase 10.1103/PhysRevLett.104.125301} {\bibfield  {journal} {\bibinfo
  {journal} {Physical Review Letters}\ }\textbf {\bibinfo {volume} {104}},\
  \bibinfo {pages} {125301} (\bibinfo {year} {2010})}\BibitemShut {NoStop}%
\bibitem [{\citenamefont {Greiner}\ \emph {et~al.}(2005)\citenamefont
  {Greiner}, \citenamefont {Regal}, \citenamefont {Stewart},\ and\
  \citenamefont {Jin}}]{Greiner2005}%
  \BibitemOpen
  \bibfield  {author} {\bibinfo {author} {\bibfnamefont {M.}~\bibnamefont
  {Greiner}}, \bibinfo {author} {\bibfnamefont {C.~A.}\ \bibnamefont {Regal}},
  \bibinfo {author} {\bibfnamefont {J.~T.}\ \bibnamefont {Stewart}}, \ and\
  \bibinfo {author} {\bibfnamefont {D.~S.}\ \bibnamefont {Jin}},\ }\href
  {\doibase 10.1103/PhysRevLett.94.110401} {\bibfield  {journal} {\bibinfo
  {journal} {Physical Review Letters}\ }\textbf {\bibinfo {volume} {94}},\
  \bibinfo {pages} {110401} (\bibinfo {year} {2005})}\BibitemShut {NoStop}%
\bibitem [{\citenamefont {F\"{o}lling}\ \emph {et~al.}(2005)\citenamefont
  {F\"{o}lling}, \citenamefont {Gerbier}, \citenamefont {Widera}, \citenamefont
  {Mandel}, \citenamefont {Gericke},\ and\ \citenamefont
  {Bloch}}]{Folling2005}%
  \BibitemOpen
  \bibfield  {author} {\bibinfo {author} {\bibfnamefont {S.}~\bibnamefont
  {F\"{o}lling}}, \bibinfo {author} {\bibfnamefont {F.}~\bibnamefont
  {Gerbier}}, \bibinfo {author} {\bibfnamefont {A.}~\bibnamefont {Widera}},
  \bibinfo {author} {\bibfnamefont {O.}~\bibnamefont {Mandel}}, \bibinfo
  {author} {\bibfnamefont {T.}~\bibnamefont {Gericke}}, \ and\ \bibinfo
  {author} {\bibfnamefont {I.}~\bibnamefont {Bloch}},\ }\href {\doibase
  10.1038/nature03500} {\bibfield  {journal} {\bibinfo  {journal} {Nature}\
  }\textbf {\bibinfo {volume} {434}},\ \bibinfo {pages} {481} (\bibinfo {year}
  {2005})}\BibitemShut {NoStop}%
\bibitem [{\citenamefont {Rom}\ \emph {et~al.}(2006)\citenamefont {Rom},
  \citenamefont {Best}, \citenamefont {van Oosten}, \citenamefont {Schneider},
  \citenamefont {F\"{o}lling}, \citenamefont {Paredes},\ and\ \citenamefont
  {Bloch}}]{Rom2006}%
  \BibitemOpen
  \bibfield  {author} {\bibinfo {author} {\bibfnamefont {T.}~\bibnamefont
  {Rom}}, \bibinfo {author} {\bibfnamefont {T.}~\bibnamefont {Best}}, \bibinfo
  {author} {\bibfnamefont {D.}~\bibnamefont {van Oosten}}, \bibinfo {author}
  {\bibfnamefont {U.}~\bibnamefont {Schneider}}, \bibinfo {author}
  {\bibfnamefont {S.}~\bibnamefont {F\"{o}lling}}, \bibinfo {author}
  {\bibfnamefont {B.}~\bibnamefont {Paredes}}, \ and\ \bibinfo {author}
  {\bibfnamefont {I.}~\bibnamefont {Bloch}},\ }\href {\doibase
  10.1038/nature05319} {\bibfield  {journal} {\bibinfo  {journal} {Nature}\
  }\textbf {\bibinfo {volume} {444}},\ \bibinfo {pages} {733} (\bibinfo {year}
  {2006})}\BibitemShut {NoStop}%
\bibitem [{\citenamefont {Spielman}\ \emph {et~al.}(2007)\citenamefont
  {Spielman}, \citenamefont {Phillips},\ and\ \citenamefont
  {Porto}}]{Spielman2007}%
  \BibitemOpen
  \bibfield  {author} {\bibinfo {author} {\bibfnamefont {I.~B.}~\bibnamefont
  {Spielman}}, \bibinfo {author} {\bibfnamefont {W.~D.}~\bibnamefont {Phillips}},
  \ and\ \bibinfo {author} {\bibfnamefont {J.~V.}~\bibnamefont {Porto}},\ }\href
  {\doibase 10.1103/PhysRevLett.98.080404} {\bibfield  {journal} {\bibinfo
  {journal} {Physical Review Letters}\ }\textbf {\bibinfo {volume} {98}},\
  \bibinfo {pages} {080404} (\bibinfo {year} {2007})}\BibitemShut {NoStop}%
\bibitem [{\citenamefont {Sogo}\ \emph {et~al.}(2009)\citenamefont {Sogo},
  \citenamefont {He}, \citenamefont {Miyakawa}, \citenamefont {Yi},
  \citenamefont {Lu},\ and\ \citenamefont {Pu}}]{Sogo2009}%
  \BibitemOpen
  \bibfield  {author} {\bibinfo {author} {\bibfnamefont {T.}~\bibnamefont
  {Sogo}}, \bibinfo {author} {\bibfnamefont {L.}~\bibnamefont {He}}, \bibinfo
  {author} {\bibfnamefont {T.}~\bibnamefont {Miyakawa}}, \bibinfo {author}
  {\bibfnamefont {S.}~\bibnamefont {Yi}}, \bibinfo {author} {\bibfnamefont
  {H.}~\bibnamefont {Lu}}, \ and\ \bibinfo {author} {\bibfnamefont
  {H.}~\bibnamefont {Pu}},\ }\href {\doibase 10.1088/1367-2630/11/5/055017}
  {\bibfield  {journal} {\bibinfo  {journal} {New Journal of Physics}\ }\textbf
  {\bibinfo {volume} {11}},\ \bibinfo {pages} {055017} (\bibinfo {year}
  {2009})}\BibitemShut {NoStop}%
\bibitem [{\citenamefont {Babadi}\ \emph {et~al.}(2013)\citenamefont {Babadi},
  \citenamefont {Skinner}, \citenamefont {Fogler},\ and\ \citenamefont
  {Demler}}]{Babadi2013}%
  \BibitemOpen
  \bibfield  {author} {\bibinfo {author} {\bibfnamefont {M.}~\bibnamefont
  {Babadi}}, \bibinfo {author} {\bibfnamefont {B.}~\bibnamefont {Skinner}},
  \bibinfo {author} {\bibfnamefont {M.~M.}\ \bibnamefont {Fogler}}, \ and\
  \bibinfo {author} {\bibfnamefont {E.}~\bibnamefont {Demler}},\ }\href
  {http://stacks.iop.org/0295-5075/103/i=1/a=16002} {\bibfield  {journal}
  {\bibinfo  {journal} {EPL (Europhysics Letters)}\ }\textbf {\bibinfo {volume}
  {103}},\ \bibinfo {pages} {16002} (\bibinfo {year} {2013})}\BibitemShut
  {NoStop}%
\bibitem [{\citenamefont {Fradkin}\ \emph {et~al.}(2010)\citenamefont
  {Fradkin}, \citenamefont {Kivelson}, \citenamefont {Lawler}, \citenamefont
  {Eisenstein},\ and\ \citenamefont {Mackenzie}}]{Fradkin2010b}%
  \BibitemOpen
  \bibfield  {author} {\bibinfo {author} {\bibfnamefont {E.}~\bibnamefont
  {Fradkin}}, \bibinfo {author} {\bibfnamefont {S.~A.}\ \bibnamefont
  {Kivelson}}, \bibinfo {author} {\bibfnamefont {M.~J.}\ \bibnamefont
  {Lawler}}, \bibinfo {author} {\bibfnamefont {J.~P.}\ \bibnamefont
  {Eisenstein}}, \ and\ \bibinfo {author} {\bibfnamefont {A.~P.}\ \bibnamefont
  {Mackenzie}},\ }\href {\doibase 10.1146/annurev-conmatphys-070909-103925}
  {\bibfield  {journal} {\bibinfo  {journal} {Annual Review of Condensed Matter
  Physics}\ }\textbf {\bibinfo {volume} {1}},\ \bibinfo {pages} {153} (\bibinfo
  {year} {2010})}\BibitemShut {NoStop}%
\bibitem [{\citenamefont {Berezinskii}(1972)}]{Berezinskii1972}%
  \BibitemOpen
  \bibfield  {author} {\bibinfo {author} {\bibfnamefont {V.~L.}\ \bibnamefont
  {Berezinskii}},\ }\href
  {http://www.jetp.ac.ru/cgi-bin/e/index/e/34/3/p610?a=list} {\bibfield
  {journal} {\bibinfo  {journal} {Sov. Phys. JETP}\ }\textbf {\bibinfo {volume}
  {34}},\ \bibinfo {pages} {610} (\bibinfo {year} {1972})}\BibitemShut
  {NoStop}%
\bibitem [{\citenamefont {Kosterlitz}\ and\ \citenamefont
  {Thouless}(1973)}]{Kosterlitz1973}%
  \BibitemOpen
  \bibfield  {author} {\bibinfo {author} {\bibfnamefont {J.~M.}\ \bibnamefont
  {Kosterlitz}}\ and\ \bibinfo {author} {\bibfnamefont {D.~J.}\ \bibnamefont
  {Thouless}},\ }\href {\doibase 10.1088/0022-3719/6/7/010} {\bibfield
  {journal} {\bibinfo  {journal} {Journal of Physics C: Solid State Physics}\
  }\textbf {\bibinfo {volume} {6}},\ \bibinfo {pages} {1181} (\bibinfo {year}
  {1973})}\BibitemShut {NoStop}%

\end{thebibliography}
\end{document}